\documentclass[%
 aip,
 amsmath,amssymb,
 reprint,%
]{revtex4-1}

\usepackage{graphicx}% Include figure files
\usepackage{dcolumn}% Align table columns on decimal point
\usepackage{bm}% bold math
\usepackage{amsmath}
\usepackage[utf8]{inputenc}
\usepackage[T1]{fontenc}
\usepackage{mathptmx}
\usepackage{etoolbox}
\usepackage{comment}
\usepackage{float}
\usepackage{appendix}
\renewcommand{\Re}{\operatorname{\rm{Re}}}

\makeatletter
\def\@email#1#2{%
 \endgroup
 \patchcmd{\titleblock@produce}
  {\frontmatter@RRAPformat}
  {\frontmatter@RRAPformat{\produce@RRAP{*#1\href{mailto:#2}{#2}}}\frontmatter@RRAPformat}
  {}{}
}%
\makeatother
\begin{document}

\preprint{AIP/123-QED}

\title[Energy exchange between electrons and ions in ion temperature gradient turbulence]{Energy exchange between electrons and ions in ion temperature gradient turbulence}
% Force line breaks with \\
\author{T. Kato}
 \email{kato.tetsuji21@ae.k.u-tokyo.ac.jp}
 \affiliation{Graduate School of Frontier Science, The University of Tokyo, Kashiwa 277-8561, Japan}

\author{H. Sugama}%
\affiliation{Graduate School of Frontier Science, The University of Tokyo, Kashiwa 277-8561, Japan}
\affiliation{National Institute for Fusion Science, Toki 509-5292, Japan}%

\author{T.-H. Watanabe}
\affiliation{Department of Physics, Nagoya University, Nagoya 464-8602, Japan}%

\author{M. Nunami}
\affiliation{National Institute for Fusion Science, Toki 509-5292, Japan}%
\affiliation{Department of Physics, Nagoya University, Nagoya 464-8602, Japan}%

\date{\today}% It is always \today, today,

%----------Abstract------------
\begin{abstract}
Microturbulence in magnetic confined plasmas contributes to energy exchange between particles of different species as well as the particle and heat fluxes. 
Although the effect of turbulent energy exchange has not been considered significant in previous studies, it is anticipated to have a greater impact than collisional energy exchange in low collisional plasmas such as those in future fusion reactors. 
In this study, gyrokinetic simulations are performed to evaluate the energy exchange due to ion temperature gradient (ITG) turbulence in a tokamak configuration.  
The energy exchange due to the ITG turbulence mainly consists of the cooling of ions in the $\nabla B$-curvature drift motion and the heating of electrons streaming along a field line.
It is found that the ITG turbulence transfers energy from ions to electrons regardless of whether the ions or electrons are hotter, which is in marked contrast to the energy transfer by Coulomb collisions. 
This implies that the ITG turbulence should be suppressed from the viewpoint of sustaining the high ion temperature required for fusion reactions since it prevents energy transfer from alpha-heated electrons to ions as well as enhancing ion heat transport toward the outside of the reactor. 
Furthermore, linear and nonlinear simulation analyses confirm the feasibility of quasilinear modeling for predicting the turbulent energy exchange in addition to the particle and heat fluxes.  
\end{abstract}

\maketitle

%----------Introduction------------
\section{Introduction}

Numerous studies on anomalous transport of particles and heat generated by microscopic turbulence have been done, based on gyrokinetic theory and simulation\cite{Antonsen, CTB, F-C, Dimits, Krommes2012, Garbet2010, Idomura2006, Horton}. 
However, there have been fewer theoretical works on energy exchange between different particle species due to turbulence,\cite{Sugama1996,Sugama2009,Hinton,Waltz} and a small number of simulations have been performed to investigate the effect of turbulent energy exchange on the evolution of temperature profiles.~\cite{Candy}
In Ref.~\cite{Candy}, it is shown from gyrokinetic simulations that the effect of turbulent energy exchange is negligibly small under conditions of DIII-D shot 128913.
However, sufficient comparative studies have not been made between collisional and turbulent energy exchanges in a wide range of conditions. 
In particular, in the case of high temperature plasmas, the impact of collisional energy exchange is expected to be small due to the low collision frequency, while turbulent energy exchange can work actively even in collisionless plasmas. 
As examples of other works related to turbulent energy exchange, there have been studies on the scaling of the turbulent transport and heating of impurities in magnetized plasmas\cite{Barnes_scaling}
and the thermal equilibration of ions and electrons by turbulence in astrophysical plasmas\cite{Kawazura}.
%In fact, there have been numerous papers on turbulent energy transfer in the space and astrophysical plasmas. Turbulence has been reported to cause thermal disequilibration of ions and electrons in collisionless plasmas\cite{Kawazura}, and the scaling of turbulent energy exchange for impurities has also been discussed\cite{Barnes_scaling}.

%
Since collisional heat transfer from alpha-heated electrons to ions is expected to play a critical role in sustaining burning plasmas in future reactors, it is an important issue to compare the effects of turbulence and collisions on the energy exchange between ions and electrons in high temperature plasmas.
In the present paper, we evaluate the energy exchange in ion temperature gradient (ITG) turbulence by gyrokinetic turbulence simulations and investigate its properties, such as dependence on the ratio between ion and electron temperatures in comparison to those of the collisional energy exchange.

To perform simulations that predict global density and temperature profiles, it is practical to use turbulent transport models such as  quasilinear ones,\cite{Casati, Bourdelle, Staebler, Narita, Toda, Parker} rather than running direct turbulence simulations for all cases. 
In fact, it has been shown that these models can reproduce particle and heat fluxes obtained from gyrokinetic turbulence simulations within acceptable errors. 
A quasilinear model for turbulent energy exchange is shown in Ref.~\cite{Barnes} where an electron drift wave instability is treated in a slab geometry with no temperature gradient. 
Detailed studies have not been conducted on modeling turbulent energy exchange under more complex conditions such as those for toroidal ITG mode remains. 
Quasilinear models are based on the assumption that the ratios of turbulent transport fluxes and energy exchange to the squared potential fluctuation amplitude estimated by linear analyses, which are called quasilinear weights\cite{Staebler}, take approximately the same values as those ratios in a steady state of turbulence obtained by nonlinear simulations. 
In this work, linear and nonlinear simulation results are compared to show the validity of this assumption in the quasilinear modeling  of energy exchange as well as particle and heat transport fluxes in the ITG turbulence. 
We here note that this work demonstrates only the feasibility of quasilinear modeling but does not present a saturation rule which is necessary for developing a quasilinear model. 
We also discuss physical mechanisms and the quasilinear modeling of turbulent energy exchange by wavenumber spectral analyses of entropy balance in microturbulence\cite{Sugama1996, Sugama2009, Nakata, Maeyama}. 

The rest of this paper is organized as follows. 
 In Secs.~\ref{subsec:2A} and \ref{subsec:2B}, gyrokinetic equations and two balance equations related to perturbed entropy density and thermal energy are presented. 
The turbulent energy exchange is represented by wavenumber spectral functions in Sec.~\ref{subsec:2C}. 
In Sec~\ref{subsec:2D}, the entropy balance in wavenumber space is investigated to consider the conditions for the quasilinear model to correctly predict the turbulent energy exchange and turbulent particle and heat transport fluxes. 
In Sec.~\ref{sec:3}, results of the ITG turbulence by the GKV code\cite{GKV}, which uses a flux tube domain\cite{Beer}, are shown. 
Simulation settings are described in Sec.~\ref{subsec:3A}, and the turbulent energy exchange and transport fluxes obtained by the GKV simulation are shown as functions of $T_e/T_i$  in Sec.~\ref{subsec:3B}. 
Comparisons between collisional and turbulent energy exchanges are made in Sec.~\ref{subsec:3C}. 
In addition, the result reported in Ref.~\cite{Candy}, where the effect of turbulent energy exchange is shown to be negligible in DIII-D shot 128913, is verified by the simulation using the same shot conditions. 
In Sec.~\ref{subsec:3D}, spectrum analyses of the turbulent energy exchange are performed to investigate its mechanisms and they are found to be directly connected with those of destabilizing the ITG modes. 
In Sec.~\ref{subsec:EB_Quasilinear}, linear and nonlinear simulation results of the entropy balance in wavenumber space are compared to confirm the validity of the assumption of the quasilinear model regarding the quasilinear weights for the turbulent transport fluxes and energy exchange.
Finally, conclusions and discussion are given in Sec.~\ref{sec:4}.

\section{Theoretical model}

\subsection{Gyrokinetic equations}\label{subsec:2A}
The plasma distribution function $F_a$ of the position vector $\bm{x}$, velocity vector $\bm{v}$, and time $t$ can be written as $F_a(\bm{x}, \bm{v}, t)=f_a(\bm{x}, \bm{v}, t)+\widehat{f}_a(\bm{x}, \bm{v}, t)$, where $f_a$ and $\widehat{f}_a$ represent the ensemble-averaged and fluctuation parts, respectively, and subscript $a$ denotes the particle species.
%[change], and $\bm{x}, \bm{v}, t$ are the position vector, velocity vector and time, respectively. 
%
The space-time scales of variations in the ensemble-averaged part $f_a$ 
are much larger than those of the fluctuation part  $\widehat{f}_a$ so that 
the ensemble average can also be regarded as the local space–time average. 
In the gyrokinetic theory, perturbations such as $\widehat{f}_a$ are assumed to satisfy the gyrokinetic ordering,
\begin{equation}
\label{eq:ordering}
     \frac{\widehat{f}_a}{f_a}\sim \frac{e_a\widehat{\phi}}{T_a} \sim \frac{|\widehat{\bm{B}}|}{|\bm{B}|} \sim \frac{\omega}{\Omega_{a}} \sim \frac{k_\parallel}{k_\perp} \sim \frac{\rho_{ta}}{L} \sim \delta \ll 1
, 
\end{equation}
where $\widehat{\phi}$ and $\widehat{\bm{B}}$ are the perturbed electrostatic potential and the perturbed magnetic field, respectively.
Here,  $\Omega_{a}=e_aB/m_ac$ , $\rho_{ta}=v_{ta}/\Omega_a$ , $m_a$, $e_a$, $T_a$, $\bm{B}$, $c$, and $v_{ta}\equiv \sqrt{T_a/m_a}$ are the gyrofrequency, thermal gyroradius, mass, charge, temperature, background magnetic field, speed of light, and thermal velocity, respectively. 
The characteristic wavenumbers (in directions parallel and perpendicular to the background magnetic field), frequency, and equilibrium scale length are represented by $k_\parallel$, $k_\perp$, $\omega$,  and $L$, respectively.

It is useful to express any perturbed function $\widehat{W}$ in the WKB form\cite{WKB},
\begin{equation}
\widehat{W}(\bm{x}, t)=\sum_{\bm{k}_\perp}{W_{\bm{k}_\perp}\exp(iS_{\bm{k}_\perp}(\bm{x}, t))},
\end{equation}
where $S_{\bm{k}_\perp}$ is the eikonal whose gradient gives the perpendicular wavenumber vector $\nabla S_{\bm{k}_\perp}=\bm{k}_\perp$. 
On the zeroth order in $\delta$,
the ensemble-averaged distribution function $f_a$ is assumed to be the local Maxwellian  
$f_{Ma}$
which is given in terms of the background density $n_a$ and temperature $T_a$ by 
\begin{equation}
f_{Ma} = 
n_a \left(\frac{m_a}{2\pi T_a} \right)^{3/2}
\exp \left( - \frac{m_a v^2}{2T_a} \right)
,
\end{equation}
and the perturbed distribution function $f_{a\bm{k}_\perp}$ can be written as
\begin{equation}
\label{eq:expression of fk}
f_{a\bm{k}_\perp}=-\frac{e_a \phi_{\bm{k}_\perp}}{T_a}f_{Ma}+ 
h_{a\bm{k}_\perp}e^{-i\bm{k}_\perp\cdot\bm{\rho}_a},
\end{equation}
where $\bm{\rho}_a=\bm{b}\times\bm{v}/\Omega_a$ with $\bm{b}=\bm{B}/B$ and $B=|\bm{B}|$. 
Here, $h_{a\bm{k}_\perp}$  represents the nonadiabatic distribution function which is calculated by the gyrokinetic equation,
\begin{eqnarray}
\label{eq:GKequation_nonadiabatic}
& &\left(\frac{\partial}{\partial t}+ v_\parallel\bm{b}\cdot\nabla+i\bm{k}_\perp\cdot(\bm{v}_E+\bm{v}_{da})\right)h_{a\bm{k}_\perp} \nonumber \\
& &=\frac{e_a}{T_a}f_{Ma}\left( \frac{\partial}{\partial t} + i\bm{k}_\perp\cdot(\bm{v}_E+\bm{v}_{*a})\right) \psi_{a\bm{k}_\perp}\nonumber \\ & &+\frac{c}{B} \sum_{\bm{k}_\perp'+\bm{k}_\perp''=\bm{k}_\perp}\left[ \bm{b}\cdot\left( \bm{k}_\perp'\times\bm{k}_\perp'' \right)\right] \psi_{a\bm{k}_\perp'}h_{a\bm{k}_\perp''}  +C_a^{GK}, 
\end{eqnarray}
where $v_\parallel=\bm{v}\cdot\bm{b}$, $\bm{v}_E=c\bm{E}\times\bm{b}/B$, $\bm{v}_{da}=c\bm{b}\times(\mu\nabla B+m_av_\parallel^2\bm{b}\cdot\nabla\bm{b})/e_a B$, $\bm{v}_{*a}= cT_a\bm{b}\times \left\{ \nabla \ln n_a + \left(w-3/2 \right)\nabla\ln T_a\right\}/e_aB$, $C_a^{GK}$ is the gyrokinetic collision term\cite{Sugama2009}, and $\bm{E}=-\nabla \Phi$ is the background electric field. 
Here, the background $E\times B$ flow is assumed to be $\bm{v}_E \sim \delta v_{ti}$, and the gradient scale length of $\bm{v}_E$ is estimated as $L$.
Therefore, the effect of electric field shear is neglected.
The detail is discussed in Appendix \ref{app: shear}.
In Eq.~(\ref{eq:GKequation_nonadiabatic}), $h_{a\bm{k}_\perp}$ is regarded as a function of time $t$ and phase space variables ($\bm{x}$, $w=m_av^2/2$, $\mu=m_av_\perp^2/2B$), where $v=|\bm{v}|$ and $v_\perp=|\bm{v}-v_\parallel \bm{b}|$. 
The gyrophase-averaged perturbed potential function $\psi_{a\bm{k}_\perp}$ is defined in terms of the perturbed electrostatic potential $\phi_{\bm{k}_\perp}$ and the perturbed vector potential $\bm{A}_{\bm{k}_\perp}$ as
\begin{eqnarray}
\label{eq: NonAdGK}
    \psi_{a\bm{k}_\perp}&=&\oint \frac{d\xi}{2\pi} \exp\left(i\bm{k}_\perp \cdot \bm{\rho}_a\right)\left( \phi_{\bm{k}_\perp}-\frac{\bm{v}}{c}\cdot\bm{A}_{\bm{k}_\perp}\right) \nonumber \\
    &=&J_0\left( k_\perp \rho_a \right)\left( \phi_{\bm{k}_\perp}-\frac{v_\parallel}{c}A_{\parallel \bm{k}_\perp} \right)+J_1\left( k_\perp \rho_a \right)\frac{v_\perp}{c}\frac{B_{\parallel \bm{k}_\perp}}{k_\perp}, \nonumber \\
\end{eqnarray}
where 
$J_0$ and $J_1$ are the zeroth- and first-order Bessel functions, respectively, $A_{\parallel \bm{k}_\perp}=\bm{b}\cdot\bm{A}_{\bm{k}_\perp}$, and $ B_{\parallel \bm{k}_\perp}=i\bm{b}\cdot ( \bm{k}_\perp\times\bm{A}_{\bm{k}_\perp}$. 
The perturbed electric and magnetic fields are determined by Poisson's equation and Ampere's law in the gyrokinetic form,
\begin{eqnarray}
(k_\perp^2+\lambda_D^{-2})\phi_{\bm{k}_\perp}=4\pi\sum_{a}{e_a\int{d^3vJ_0(k_\perp \rho_a)h_{a\bm{k}_\perp}}} \label{eq:Poisson}\\
k_\perp^2A_{\parallel \bm{k}_\perp}=\frac{4\pi}{c}\sum_{a}{e_a\int{d^3vJ_0(k_\perp \rho_a)h_{a\bm{k}_\perp}}}v_\parallel \label{eq:A_para}\\
-k_\perp B_{\parallel \bm{k}_\perp}=\frac{4\pi}{c}\sum_{a}{e_a\int{d^3vJ_1(k_\perp \rho_a)h_{a\bm{k}_\perp}}v_\perp} \label{eq:B_para}
,
\end{eqnarray}
where $\lambda_D=1/\sqrt{\sum_a{4\pi n_ae_a^2/T_a}}$ is the Debye length. 
The time evolution of nonadiabatic distribution function $h_{a\bm{k}_\perp}$ can be obtained by 
solving Eq.(\ref{eq:GKequation_nonadiabatic}) combined with Eqs.~(\ref{eq:Poisson})--(\ref{eq:B_para}).

\subsection{Energy and entropy balance equations}\label{subsec:2B}
We now consider two balance equations related to the perturbed entropy density and the energy. 
These variables are associated with the particle and heat fluxes and the energy exchange between electrons and ions. 
The entropy density due to turbulence $\delta S_a$ is defined by the difference between the macroscopic entropy density $S_{Ma}\equiv - \int d^3v f_{a}\log f_{a}$ and the ensemble averaged microscopic entropy density $\langle S_{ma}\rangle_{\rm ens}\equiv-\left\langle \int d^3v\left( f_{a}+\widehat{f}_{a}\right) \log \left( f_{a}+\widehat{f}_{a}\right)\right\rangle_{\rm ens}$ as 
\begin{equation}
\label{eq: EntropyDencity}
\delta S_a\equiv S_{Ma} 
- \langle S_{ma}\rangle_{\rm ens} 
=\sum_{\bm{k}_\perp} \left\langle \int d^3v \frac{\left| f_{a\bm{k}_\perp} \right|^2}{2f_{Ma}}\right\rangle_{\rm ens}, 
\end{equation}
where $\langle \cdots \rangle_{\rm ens}$ represents the ensemble average, and terms of $\mathcal{O}(\delta^3)$ are neglected. 
Using Eq. (\ref{eq:expression of fk}), the integral in the angle bracket in Eq. (\ref{eq: EntropyDencity}) can be rewritten as 
\begin{equation}
\int d^3v \frac{\left| f_{a\bm{k}_\perp} \right|^2}{2f_{Ma}} =\int d^3v \frac{\left| h_{a\bm{k}_\perp} \right|^2}{2f_{Ma}} -\frac{n_ae_a^2}{2T_a^2}\left|\phi_{\bm{k}_\perp}\right|^2-\frac{e_a}{T_a}\rm{Re}\left[\it{ \phi_{\bm{k}_\perp}^* n_{a\bm{k}_\perp}} \right] 
,
\end{equation}
where $*$ denotes the complex conjugate and 
$n_{a\bm{k}_\perp}\equiv\int{d^3v f_{a\bm{k}_\perp}}$ is the density perturbation with 
the perpendicular wavenumber vector  $\bm{k}_\perp$.
Multiplying Eq. (\ref{eq:GKequation_nonadiabatic}) by $h_{a\bm{k}_\perp}^*/f_{Ma}$ and taking the ensemble and flux-surface averages, we can derive the entropy balance equation,
%\frac{\partial}{\partial t}\left(\sum_{\bm{k_\perp}}dS_{\bm{k}_\perp}\right) = 
%
\begin{eqnarray}
\label{eq:EBequation_h}
\frac{\partial}{\partial t}\left(\delta S_{ha}\right)
&=& \frac{\Gamma^{\rm turb}_a}{L_{pa}}+\frac{q^{\rm turb}_a}{T_aL_{Ta}}+\frac{Q^{\rm turb}_a}{T_a}+D_a
%\\ &=& \sum_{i}{\mathcal{J}_{is}\mathcal{A}_{is}}+D_s
\end{eqnarray}
where $\delta S_{ha}=\sum_{\bm{k_\perp}} \Big\langle\Big\langle \int d^3v \left| h_{a\bm{k}_\perp} \right|^2/2f_{Ma} \Big\rangle\Big\rangle$ and $\langle \langle \cdots \rangle \rangle$ denotes a double average over the ensemble and the flux surface. 
The turbulent particle and heat fluxes $\Gamma_a^{\rm turb}, q_a^{\rm turb}$ and the inverse gradient scale lengths $L_{pa}^{-1}, L_{Ta}^{-1}$ are defined by
\begin{eqnarray}
    &&\left[\Gamma^{\rm turb}_{a}, \frac{q^{\rm turb}_{a}}{T_a}\right] 
    \equiv \sum_{\bm{k}_{\perp}} \left[ \Gamma^{\mathrm{turb}}_{a \bm{k}_{\perp}},  \frac{q^{\mathrm{turb}}_{a \bm{k}_{\perp}}}{T_a} \right] \nonumber \\ 
    &&=\sum_{\bm{k}_\perp}\Re\Bigg\langle\Bigg\langle \int d^3v \left[ 1, \frac{w}{T_a}-\frac{5}{2}\right]\nonumber\\
    && \hspace{2cm}\times h_{a\bm{k}_\perp}^*\left(-i\frac{c}{B}\psi_{a\bm{k}_\perp}\bm{k}_\perp\times\bm{b}\right)\cdot\nabla r \Bigg\rangle\Bigg\rangle \label{eq:particle and heat fluxes}\\
    &&\left[ L_{pa}^{-1}, L_{Ta}^{-1}\right]=\left[ -\frac{\partial \ln p_a}{\partial r}-\frac{e_a}{T_a}\frac{\partial\Phi}{\partial r}, -\frac{\partial \ln T_a}{\partial r}\right]\label{eq:gradients}
\end{eqnarray}
respectively, where the minor radius $r$ is used as a label of flux surfaces of a toroidal plasma and $Q_a^{\rm turb}$ represents the turbulent heating of the particles of species $a$ given by
\begin{equation}
\label{eq:TurbulentEnergyExchange}
    Q^{\rm turb}_a\equiv \sum_{\bm{k}_{\perp}} Q^{\rm turb}_{a\bm{k}_\perp}=e_a\sum_{\bm{k}_\perp}\Re\Bigg\langle\Bigg\langle \int d^3v h_{a\bm{k}_\perp}^*\frac{\partial\psi_{a\bm{k}_\perp}}{\partial t} \Bigg\rangle\Bigg\rangle
.
\end{equation}
Taking the summation of Eq.(\ref{eq:TurbulentEnergyExchange}) over the particle species and using Eqs.(\ref{eq:Poisson}), (\ref{eq:A_para}), and (\ref{eq:B_para}), we obtain 
\begin{eqnarray}
 & &   \sum_{a}{Q^{\rm turb}_a}
\begin{comment}
=
\sum_a 
\sum_{\bm{k}_\perp} \frac{e_a}{2}
\frac{\partial}{\partial t} 
\Re\Bigg\langle\Bigg\langle \int d^3v h_{a\bm{k}_\perp}^* \psi_{a\bm{k}_\perp} \Bigg\rangle\Bigg\rangle
\end{comment}
 =
\frac{1}{8\pi}\frac{\partial}{\partial t} \sum_{\bm{k}_\perp}\Bigg\langle\Bigg\langle \left( k_\perp^2+\lambda_D^{-2} \right)|\phi_{\bm{k}_\perp}|^2 - |\bm{B}_{\bm{k}_\perp}|^2 \Bigg\rangle\Bigg\rangle,\nonumber \\
\end{eqnarray}
where $\bm{B}_{\bm{k}_\perp}=i \bm{k}_\perp\times\bm{A}_{\bm{k}_\perp}$.
Using the transport time scale ordering $\left( \partial \langle \langle \cdots \rangle \rangle / \partial t\right)/\langle \langle \cdots \rangle \rangle = \mathcal{O}(\delta^2)$, we obtain $\sum_{a}{Q^{\rm turb}_a}=0$ to the lowest order in $\delta$. Therefore, we can interpret $Q^{\rm turb}_a$ as the energy exchange between different plasma species. 
The last term in Eq. (\ref{eq:EBequation_h}) $D_a$ represents collisional dissipation written as
\begin{equation}
%\label{eq:TurbulentEnergyExchange}
    D_a\equiv\sum_{\bm{k}_{\perp}} D_{a\bm{k}_\perp}=\sum_{\bm{k}_\perp}\Re\Bigg\langle\Bigg\langle \int d^3v \frac{ h_{a\bm{k}_\perp}^*}{f_{Ma}}C_a^{GK} \Bigg\rangle\Bigg\rangle
\end{equation}
The left-hand side of Eq.~(\ref{eq:EBequation_h}) vanishes in the steady state of turbulence 
where the entropy production due to turbulent transport and heating balance with 
the collisional dissipation. 
%
%Next, the role of turbulence on the flux-surface averaged energy balance equation is considered. 
%One can derive particle and energy conservation laws with turbulence effects from Vlasov equation including the second order perturbed term. 
%The detail derivation relies on reference. 
%As a result, the energy balance equations in the axisymmetric systems on the transport time scale are shown as
%

The ensemble and flux-surface averaged energy balance equation is given by\cite{Sugama1996}
\begin{eqnarray}
% \label{eq:particle conservation}
%& &\frac{\partial n_s}{\partial t} +\nabla\cdot \Gamma_s=0 \\
%\label{eq:transport equation}
\label{eq:transport equation}
& &\frac{\partial}{\partial t}\left(\frac{3}{2}p_a \right) +
\frac{1}{V'} \frac{\partial }{\partial r}
\left[ V' \left(q_a+\frac{5}{2}T_a\Gamma_a \right) \right]  
\nonumber \\
& & 
%\hspace{12pt} 
= \frac{( \Gamma_a^{\rm{cl}}+\Gamma_a^{\rm{ncl}})}{n_a}
\frac{\partial p_a}{\partial r}
+
\langle \bm{u}_a \cdot ( \nabla \cdot \bm{\pi}_a  ) \rangle 
+ Q^{\rm coll}_a 
\nonumber \\
& & 
\hspace*{5mm}
%\mbo{}
- e_a \Gamma_a^{\rm{turb}} \frac{\partial \Phi}{\partial r}
+ Q^{\rm turb}_a
 \end{eqnarray}
where $\langle \cdots \rangle$ represents the flux-surface average, and $p_a$, $\bm{\pi}_a$ are the pressure and viscosity tensor, respectively. 
Here, $V^\prime$ is the derivative of the volume $V$ inside a magnetic surface with respect to the minor radius $r$. 
The radial particle and heat fluxes denoted by $\Gamma_a$ and $q_a$, respectively, are written as
 \begin{eqnarray}
    & &\Gamma_a=\Gamma_a^{\rm{cl}}+\Gamma_a^{\rm{ncl}}+\Gamma_a^{\rm{turb}}\\
    & &q_a=q_a^{\rm{cl}}+q_a^{\rm{ncl}}+q_a^{\rm{turb}}
\end{eqnarray}
where the superscripts ${\rm cl}$, ${\rm nc}$, and ${\rm turb}$ denote classical, neoclassical and turbulence parts, respectively. 
Here, $Q^{\rm coll}_a$ is the flux-surface average of the collisional heat generation rate. 
In the case of a plasma consisting of electrons and single-species ions, we have
\begin{eqnarray}
\label{eq:Collisional energy exchange}
    Q^{\rm coll}_i=\frac{3m_e}{m_i}n_e \nu_e\left( T_e-T_i\right),\\
    \nu_e=\frac{4\sqrt{2\pi} n_ie_e^2e_i^2\ln \Lambda}{3m_e^{\frac{1}{2}}T_e^{\frac{3}{2}}},
\end{eqnarray}
and 
\begin{equation}
\label{eq:Collisional energy exchange_electron}
    Q^{\rm coll}_e  
=  
- \langle {\bf R}_e \cdot ( {\bf u}_e - {\bf u}_i ) \rangle_a
  - Q^{\rm coll}_i  
\end{equation}
where $\nu_e$ is the electron-ion collision frequency, $\ln \Lambda$ is the Coulomb logarithm, ${\bf R}_e$ the collisional friction force for electrons, 
and $( {\bf u}_e - {\bf u}_i )$ the difference between the electron and ion flow velocities. 
We see that the collisional energy transfer $Q^{\rm coll}_i$ from electrons to ions is proportional to the product of the collision frequency and the temperature difference between electrons and ions. 

Appendix \ref{app: shear} shows that in the present model, the effect of the radial electric field $E_r$ on turbulent energy exchange appears only in the Doppler shift.
The sum of the last two terms on the right-hand side
of Eq.~(\ref{eq:transport equation}) is essentially equivalent to turbulent energy exchange Eq.~(\ref{eq:TurbulentEnergyExchange}) in $E_r=0$.
Therefore, we henceforth discuss turbulent energy exchange with $E_r=0$.

Electrons and ions exchange energy via collisions and turbulence. 
The collisional energy exchange decreases for low collision frequency. 
It always transfers energy from the hotter species to the colder one and 
vanishes when the two species have the same temperature. 
As shown later from the gyrokinetic analysis and simulation, 
the turbulent energy exchange has quite different properties from 
the collisional one. 

\subsection{Spectral analysis of turbulent energy exchange in wavenumber space}\label{subsec:2C}
%
\begin{comment}
 
\begin{eqnarray}
& & 
-
\sum_a e_a \sum_{\bm{k}_\perp}\Re\Bigg\langle\Bigg\langle \int d^3v \psi_{a \bm{k}_\perp}^*\frac{\partial h_{a\bm{k}_\perp}}{\partial t} \Bigg\rangle\Bigg\rangle 
\nonumber \\ 
& &
= 
- \frac{1}{8\pi}
\frac{\partial }{\partial t}
\left(|{\bf E}_{\bm{k}_\perp}|^2 - |{\bf B}_{\bm{k}_\perp}|^2
+ \frac{|\phi_{\bm{k}_\perp}|^2}{\lambda_D^2}
\right) 
\end{eqnarray}
%   
\end{comment}

We investigate the physical mechanism of the turbulent energy exchange by using Eq. (\ref{eq:TurbulentEnergyExchange}). 
In the steady state of turbulence, the time derivative acting on the perturbed potential Eq. (\ref{eq:GKequation_nonadiabatic}) can be transferred to that on the nonadiabatic perturbed distribution function with the help of the Leibniz rule. 
As shown in Appendix \ref{app: derivation}, using the gyrokinetic equation, Eq. (\ref{eq:GKequation_nonadiabatic}),  we can rewrite Eq. (\ref{eq:TurbulentEnergyExchange}) as
\begin{eqnarray}
\label{eq:TurbulentEnergyExchange_PowerFlow}
Q^{\rm turb}_{a}&=&-e_a\sum_{\bm{k}_\perp}\Re\Bigg\langle\Bigg\langle \int d^3v \psi_{a\bm{k}_\perp}^*\frac{\partial h_{a\bm{k}_\perp}}{\partial t} \Bigg\rangle\Bigg\rangle \nonumber \\
&=&\sum_{\bm{k}_\perp}\left( Q^{\rm turb}_{a\parallel\bm{k}_\perp}+Q^{\rm turb}_{aB\bm{k}_\perp}+Q^{\rm turb}_{a\psi\bm{k}_\perp}+Q^{\rm turb}_{aC\bm{k}_\perp}\right) \label{eq:TEE_parts}
\end{eqnarray}
with
\begin{eqnarray}
Q^{\rm turb}_{a\parallel\bm{k}_\perp}&=&\Re\Bigg\langle\Bigg\langle \int d^3v \bm{G}_{a\parallel\bm{k}_\perp}\cdot\bm{j}_{a\parallel\bm{k}_\perp}^* \Bigg\rangle\Bigg\rangle \label{eq: parallel_heating}\\
Q^{\rm turb}_{aB\bm{k}_\perp}&=&\Re\Bigg\langle\Bigg\langle \int d^3v 
 \bm{G}_{a\perp\bm{k}_\perp}\cdot\bm{j}_{aB\bm{k}_\perp}^* \Bigg\rangle\Bigg\rangle 
 \label{eq:perp_heating}\\
Q^{\rm turb}_{a\psi\bm{k}_\perp}&=&\Re\Bigg\langle\Bigg\langle \int d^3v \bm{G}_{a\perp\bm{k}_\perp}\cdot \bm{j}_{a\psi\bm{k}_\perp}^* \Bigg\rangle\Bigg\rangle 
\label{eq:psi_heating}\\
Q^{\rm turb}_{aC\bm{k}_\perp}&=&-\Re\Bigg\langle\Bigg\langle \int d^3v C_a^{GK} \psi_{a\bm{k}_\perp}^* \Bigg\rangle\Bigg\rangle 
\label{eq:collision_psi}
\end{eqnarray}
where $E_r=0$ is used. The perturbed fields $\bm{G}_{a\bm{k}_\perp}$ and
the perturbed currents $\bm{j}_{a\bm{k}_\perp}$ at the gyrocenter position are 
defined by
%
\begin{comment}
\begin{eqnarray}
\bm{G}_{a\bm{k}_\perp} 
& = & -\nabla_\parallel \psi_{a\bm{k}_\perp}-i\bm{k}_\perp\psi_{a\bm{k}_\perp}
=\bm{G}_{a\parallel\bm{k}_\perp }+\bm{G}_{a\perp\bm{k}_\perp } , \label{eq:Gfield}
\\ 
\bm{j}_{a\bm{k}_\perp} & = & 
e_a \left( \bm{u}_{a\parallel\bm{k}_\perp}+\bm{u}_{a\perp\bm{k}_\perp} \right) 
=\bm{j}_{a\parallel\bm{k}_\perp }+\bm{j}_{a\perp\bm{k}_\perp }, 
\\
\bm{u}_{a\parallel\bm{k}_\perp} 
& = &
h_{a\bm{k}_\perp}v_\parallel\bm{b}, 
\\ \bm{u}_{a\perp\bm{k}_\perp}
& = & \bm{u}_{aB\bm{k}_\perp}+\bm{u}_{a\psi\bm{k}_\perp}, 
\\ 
\bm{u}_{aB\bm{k}_\perp} 
& = & 
h_{a\bm{k}_\perp}\bm{v}_{sd}, 
\\ 
\bm{u}_{a\psi\bm{k}_\perp}  &= & 
\frac{ic}{B} \sum_{\bm{k}_\perp'+\bm{k}_\perp''= 
\bm{k}_\perp}\left( \bm{b}\times\bm{k}_\perp'\right) \psi_{a\bm{k}_\perp'}h_{a\bm{k}_\perp''} .\label{eq:u_psi}
\end{eqnarray}
\end{comment}
%
\begin{eqnarray}
\bm{G}_{a\bm{k}_\perp} 
& = & -\nabla_\parallel \psi_{a\bm{k}_\perp}-i\bm{k}_\perp\psi_{a\bm{k}_\perp}
=\bm{G}_{a\parallel\bm{k}_\perp }+\bm{G}_{a\perp\bm{k}_\perp } , \label{eq:Gfield}
\\ 
\bm{j}_{a\bm{k}_\perp} & = & 
\bm{j}_{a\parallel\bm{k}_\perp }+\bm{j}_{a\perp\bm{k}_\perp }, 
\\
\bm{j}_{a\parallel\bm{k}_\perp} 
& = &
e_ah_{a\bm{k}_\perp}v_\parallel\bm{b}, 
\\ \bm{j}_{a\perp\bm{k}_\perp}
& = & \bm{j}_{aB\bm{k}_\perp}+\bm{j}_{a\psi\bm{k}_\perp}, 
\\ 
\bm{j}_{aB\bm{k}_\perp} 
& = & 
e_ah_{a\bm{k}_\perp}\bm{v}_{da}, 
\\ 
\bm{j}_{a\psi\bm{k}_\perp}  &= & 
\frac{ice_a}{B} \sum_{\bm{k}_\perp'+\bm{k}_\perp''= 
\bm{k}_\perp}\left( \bm{b}\times\bm{k}_\perp'\right) \psi_{a\bm{k}_\perp'}h_{a\bm{k}_\perp''} .\label{eq:u_psi}
\end{eqnarray}
Here,
$\bm{j}_{a\psi\bm{k}_\perp}$ is derived from the nonlinear term of Eq. (\ref{eq:GKequation_nonadiabatic}), and represents the current induced by the perturbed potential. 
As shown in Eq. (\ref{eq:TurbulentEnergyExchange_PowerFlow}), the turbulent energy exchange is caused by the product of the perturbed field and current due to the nonadiabatic distribution. 
The field and current can be decomposed into components in directions parallel and perpendicular to the background magnetic field. 
The perpendicular current component can also be classified as the two parts: 
the one is  produced by the $\nabla B$-curvature drift in the toroidal magnetic field and the other by the drift due to the turbulent potential field. 
Furthermore,  the effect of collisions is given by the last term at the right-hand side of Eq.~(\ref{eq:TurbulentEnergyExchange_PowerFlow}), and thus the turbulent energy exchange is represented by the sum of the four parts.  

As mentioned earlier, in the steady state of turbulence, the turbulent energy exchange does not cause a net increase or decrease in energy. 
This property is also valid for each wavenumber as shown by 
 \begin{equation}
 \label{eq: no net heating}
    \sum_a Q^{\rm turb}_{a\bm{k}_\perp}=0
 \end{equation}
where $Q^{\rm turb}_{a\bm{k}_\perp}=Q^{\rm turb}_{a\parallel\bm{k}_\perp} +Q^{\rm turb}_{aB\bm{k}_\perp}+Q^{\rm turb}_{a\psi\bm{k}_\perp}+Q^{\rm turb}_{aC\bm{k}_\perp}$. 
Furthermore,  we have
 \begin{equation}
\label{sumqpsi}
    \sum_{\bm{k}_\perp} Q^{\rm turb}_{a\psi\bm{k}_\perp}=0
 \end{equation}
which can be illustrated by using the definition of $Q^{\rm turb}_{a\psi\bm{k}_\perp}$ given in
Eqs.~(\ref{eq:psi_heating}), (\ref{eq:Gfield}) and (\ref{eq:u_psi}). 
Equation~(\ref{sumqpsi}) implies that $Q^{\rm turb}_{a\psi\bm{k}_\perp}$ does not contribute to the net heating or cooling of  particles of the species $a$ but it represents the energy transfer in the wavenumber space through nonlinear interactions between different modes. 
Thus, $Q^{\rm turb}_{a\psi\bm{k}_\perp}$ influences the profile of the total wavenumber spectrum $Q^{\rm turb}_{a\bm{k}_\perp}$. 
Because of Eq.~(\ref{sumqpsi}), the total turbulent energy exchange 
$Q^{\rm turb}_a$
is given by taking the sum of the three components $Q^{\rm turb}_{a\parallel\bm{k}_\perp}$, $Q^{\rm turb}_{aB\bm{k}_\perp}$, and $Q^{\rm turb}_{aC\bm{k}_\perp}$ over the whole wavenumber space. 
In particular, $Q^{\rm turb}_{a\parallel\bm{k}_\perp}$, $Q^{\rm turb}_{aB\bm{k}_\perp}$ are the contributions to Joule heating (cooling) via currents parallel and perpendicular to the background magnetic field.

It is noted here that, 
using Eq.~(\ref{eq:GKequation_nonadiabatic}) and Eqs.~(\ref{eq:Poisson})--(\ref{eq:B_para}),  
we can obtain
\begin{eqnarray}
\label{psidth}
& & 
-
\sum_a e_a \sum_{\bm{k}_\perp}\Re\Bigg\langle\Bigg\langle \int d^3v \psi_{a \bm{k}_\perp}^*\frac{\partial h_{a\bm{k}_\perp}}{\partial t} \Bigg\rangle\Bigg\rangle 
\nonumber \\ 
& &
= 
- 
\frac{\partial }{\partial t}
\sum_{\bm{k}_\perp}
\left[ \frac{1}{8\pi}
\Big\langle \Big\langle
|{\bf E}_{\bm{k}_\perp}|^2 - |{\bf B}_{\bm{k}_\perp}|^2
\Big\rangle \Big\rangle
+ \sum_a\frac{n_a e_a^2\langle\langle|\phi_{\bm{k}_\perp}|^2\rangle \rangle}{2T_a}
\right]
, \nonumber \\
\end{eqnarray}
where 
the steady state of turbulence is not assumed. 
Under an electrostatic approximation, magnetic fluctuations are neglected, and 
the right-hand side of Eq.~(\ref{psidth}) is regarded as the decrease rate of 
the energy associated with the electrostatic electric field. 
Then, it is understood from Eqs.~(\ref{eq:TurbulentEnergyExchange_PowerFlow}) 
and (\ref{psidth}) that 
the turbulent heating of particles represented by 
$
\sum_a \sum_{\bm{k}_\perp} 
(
Q^{\rm turb}_{a\parallel\bm{k}_\perp} 
+ Q^{\rm turb}_{aB\bm{k}_\perp}
+ Q^{\rm turb}_{aC\bm{k}_\perp}
)
$
is brought about by consuming the electrostatic electric field energy.

\subsection{Quasilinear model and entropy balance}\label{subsec:2D}
In this section, the predictability of the quasilinear model for the turbulent energy 
exchange is discussed with the entropy balance in the linear and nonlinear states. 
When using the solutions $h_{a\bm{k}_\perp}$ and $\psi_{a\bm{k}_\perp}$ of 
the linearized version of Eq.~(\ref{eq:GKequation_nonadiabatic}) combined with Eqs.~(\ref{eq:Poisson})--(\ref{eq:B_para}) 
to evaluate $Q^{\rm turb}_a$ defined in Eq.~(\ref{eq:TurbulentEnergyExchange}), 
we obtain
\begin{equation}
\label{QL-Qturb}
    Q^{\rm turb}_a = e_a \sum_{\bm{k}_\perp}  \Re 
\biggl\langle
\biggl\langle
\int d^3v 
(- i \omega_{r \bm{k}_\perp} + \gamma_{\bm{k}_\perp} )
h^*_{a\bm{k}_\perp}\psi_{a\bm{k}_\perp}
\biggr\rangle 
\biggr\rangle 
,
\end{equation}
where $\omega_{r \bm{k}_\perp}$ and $\gamma_{\bm{k}_\perp}$ are 
the real and imaginary parts of the complex-valued linear eigenfrequency 
$\omega_{\bm{k}_\perp} \equiv \omega_{r \bm{k}_\perp} + i \gamma_{\bm{k}_\perp}$
for the wavenumber vector $\bm{k}_\perp$. 
Although $\sum_a Q^{\rm turb}_a = 0$ holds in the steady state of turbulence, 
it does not for the linear solutions. 
The presence of the finite growth rate $\gamma_{\bm{k}_\perp}$ causes 
$\sum_a Q^{\rm turb}_a \neq 0$. 
Then, we drop $\gamma_{\bm{k}_\perp}$ from Eq.~(\ref{QL-Qturb}) 
and define 
\begin{equation}
    Y_a = e_a \sum_{\bm{k}_\perp}  \Re 
\biggl\langle
\biggl\langle
\int d^3v 
(- i \omega_{r \bm{k}_\perp} )
h^*_{a\bm{k}_\perp}\psi_{a\bm{k}_\perp}
\biggr\rangle 
\biggr\rangle 
,
\end{equation}
to estimate the turbulent energy exchange from the linear solutions  
because $\sum_a Y_a = 0$ is satisfied. 
We now note that $Y_a$ can be rewritten as 
\begin{eqnarray}
    Y_a &\equiv& \sum_{\bm{k}_\perp} Y_{a\bm{k}_\perp}\nonumber\\
    &=&\frac{e_a}{2} \sum_{\bm{k}_\perp}
\Re\Bigg\langle\Bigg\langle \int d^3v 
\left(h_{a\bm{k}_\perp}^*\frac{\partial\psi_{a\bm{k}_\perp}}{\partial t} 
- \frac{\partial h^*_{a\bm{k}_\perp}}{\partial t} \psi_{a\bm{k}_\perp}\right) 
\Bigg\rangle\Bigg\rangle,\nonumber\\
    \label{eq:Candy}
\end{eqnarray}
which is the same as 
the turbulent energy exchange introduced by Candy\cite{Candy}. 
This expression of $Y_a$ in Eq.~(\ref{eq:Candy}) 
can be used for both linear and nonlinear cases 
and rigorously satisfies $\sum_a Y_a = 0$ even in non-steady states. 
The relation between $Q^{\rm turb}_a$ and $Y_a$ is given from 
Eqs.~(\ref{eq:TurbulentEnergyExchange}) and (\ref{eq:Candy}) as 
\begin{equation}
    Q^{\rm turb}_a-Y_a=\frac{\partial}{\partial t}\left\{\frac{e_a}{2} \sum_{\bm{k}_\perp}\Re\Bigg\langle\Bigg\langle \int d^3v h^*_{a \bm{k}_\perp}\psi_{a \bm{k}_\perp}\Bigg\rangle\Bigg\rangle\right\}
    \label{eq:difference}
, 
\end{equation}
from which we easily see that $Q^{\rm turb}_a = Y_a$ in the turbulent steady state. 
The entropy balance equation for each wavenumber is derived from Eq.~(\ref{eq:GKequation_nonadiabatic}) as 
\begin{eqnarray}
\frac{\partial}{\partial t}
\Bigg\langle\Bigg\langle \int d^3v \frac{\left| h_{a\bm{k}_\perp} \right|^2}{2f_{Ma}}\Bigg\rangle\Bigg\rangle   
& = & \frac{\Gamma^{\rm turb}_{a\bm{k}_\perp}}{L_{pa}}+\frac{q^{\rm turb}_{a\bm{k}_\perp}}{T_aL_{Ta}}+\frac{Q^{\rm turb}_{a \bm{k}_\perp}}{T_a}
\nonumber \\ &  & \mbox{} 
+ D_{a\bm{k}_\perp}
+ N_{a\bm{k}_\perp}  
,
\label{eq:h_wavenumber}
\end{eqnarray}
where
the subscript $\bm{k}_\perp$ denotes the contribution from each perpendicular 
wavenumber vector and 
$N_{a\bm{k}_\perp}$ is defined by
\begin{eqnarray}
&&N_{a\bm{k}_\perp}=\nonumber \\
&& \Bigg\langle\Bigg\langle\int d^3v \frac{c}{Bf_{Ma}} 
\sum_{\bm{k}_\perp'+\bm{k}_\perp''=\bm{k}_\perp}
\left[ \bm{b}\cdot\left( \bm{k}_\perp'\times\bm{k}_\perp'' \right)\right] 
\psi_{a\bm{k}_\perp'}h_{a\bm{k}_\perp''}h_{a\bm{k}_\perp}^*\Bigg\rangle\Bigg\rangle.\nonumber \\
\label{eq:entropytransfer}
\end{eqnarray}
Here, $N_{a\bm{k}_\perp}$ represents the entropy which the mode with 
the wavenumber vector $\bm{k}_\perp$ gains through nonlinear interaction
with other modes, 
and it satisfies
\begin{equation}
\sum_{\bm{k}_\perp} N_{a\bm{k}_\perp}=0
,
\label{eq:entropytransfer_2}
\end{equation}
which implies that the nonlinear interaction produces no net entropy. 

Substituting Eq.~(\ref{eq:difference}) into Eq.~(\ref{eq:h_wavenumber}) and using the
following relation, 
\begin{eqnarray}
\label{eq:entropy_and_energy}
&&\Bigg\langle\Bigg\langle \int d^3v 
\Bigg[
\frac{\left| h_{a\bm{k}_\perp} \right|^2}{2f_{Ma}} 
- \frac{e_a}{2T_a} \Re \left[ h_{a\bm{k}_\perp}^*\psi_{a\bm{k}_\perp}\right]
\Bigg]\Bigg\rangle\Bigg\rangle  
\nonumber \\
& &
= 
\Bigg\langle\Bigg\langle \int d^3v 
\frac{| f_{a\bm{k}_\perp} |^2}{2f_{Ma}}
+ \frac{e_a}{2 T_a} \bigg( n_{a \bm{k}_\perp}^*  \phi_{\bm{k}_\perp}
+ \frac{1}{c} n_a 
\bm{u}_{a \bm{k}_\perp}^* \cdot A_{\bm{k}_\perp}
\bigg)             
\Bigg\rangle\Bigg\rangle  
,
\nonumber \\
& & 
\end{eqnarray}
we obtain 
\begin{eqnarray}
&&
\frac{\partial}{\partial t} 
\Bigg\langle\Bigg\langle \int d^3v 
\frac{| f_{a\bm{k}_\perp} |^2}{2f_{Ma}}
+ \frac{e_a}{2 T_a} \bigg( n_{a \bm{k}_\perp}^*  \phi_{\bm{k}_\perp}
+ \frac{1}{c} n_a 
\bm{u}_{a \bm{k}_\perp}^* \cdot A_{\bm{k}_\perp}
\bigg)             
\Bigg\rangle\Bigg\rangle  
\nonumber \\
& & 
\hspace*{5mm} - N_{a\bm{k}_\perp} - D_{a\bm{k}_\perp}
\nonumber \\
&  & 
\hspace*{3mm}
= 
 \frac{\Gamma^{\rm turb}_{a\bm{k}_\perp}}{L_{pa}}+\frac{q^{\rm turb}_{a\bm{k}_\perp}}{T_a L_{Ta}}+\frac{Y_{a\bm{k}_\perp}}{T_a}
,
\label{eq:EBE_wavenumber}
\end{eqnarray}
where $\bm{u}_{a \bm{k}_\perp}$ is the 
perturbed flow velocity with the perpendicular wavenumber
vector $\bm{k}_\perp$ defined by 
$n_a 
\bm{u}_{a \bm{k}_\perp}
\equiv
\int d^3 v \; f_{a\bm{k}_\perp} {\bf v}$.  
In the double angle brackets on the left-hand side of 
Eq.~(\ref{eq:EBE_wavenumber}), we find  
the entropy due to the perturbed particle distribution function 
$f_{a\bm{k}_\perp}$
as well as  the inverse temperature multiplied by 
the electrostatic potential energy and the interaction 
between the magnetic potential and the current of particle species $a$.

In the quasilinear model, 
the transport fluxes divided by the squared potential, 
$\Gamma^{\rm turb}_{a\bm{k}_\perp}/ \langle \langle |\phi_{\bm{k}_\perp}|^2 \rangle \rangle$ 
and 
$q^{\rm turb}_{a\bm{k}_\perp}/
\langle \langle |\phi_{\bm{k}_\perp}|^2 \rangle \rangle$,  
in the turbulent steady state are approximated by 
the corresponding values obtained from the linear analysis. 
We here include the turbulent energy exchange term into 
the quasilinear model and estimate 
$Y_{a\bm{k}_\perp}/
\langle \langle |\phi_{\bm{k}_\perp}|^2 \rangle \rangle$ 
from the linear calculation. 
\begin{table*}[t]
\begin{minipage}[t]{0.6\textwidth}
\centering
    \caption{Plasma and field parameters}
\begin{tabular}{lcc} \hline
     \multicolumn{2}{c}{Plasma and field parameters}&   Value \\ \hline
    Normalized ion temperature gradient & $R_0/L_{Ti}$  &   6.92    \\
    Normalized electron temperature gradient & $R_0/L_{Te}$  &   0    \\
    Normalized density gradient& $R_0/L_{na}$  &   2.22    \\
    Mass ratio & $m_e/m_i$  &   $5.446\times10^{-4}$    \\
    Charge ratio& $e_e/e_i$  &   -1    \\
    Ion beta value& $\beta_i= 4 \pi n_iT_i/B^2$  &   $1\times10^{-4}$    \\
    Normalized collision frequency& $\nu^*_{ii}\equiv R_0 q_0 \nu_{ii}/(\epsilon_r^{3/2}v_{ti})$  &   0.068    \\
    Temperature ratio& $T_e/T_i$  &   $0.8 \sim 1.5$    \\ 
    Inverse aspect ratio& $\varepsilon_r$  &   $0.18$    \\ 
    Safety factor & $q_0$ & $1.4$ \\
    Magnetic shear & $\hat{s}=(r/q)(dq/dr)$ & $0.78$ \\
    \hline
    \label{tab:Plasma and field parameters}
\end{tabular}
\end{minipage}
\begin{minipage}[t]{0.35\textwidth}
    \centering
    \caption{Resolution settings}
\begin{tabular}{c} \hline
     Domain sizes and \\ resolved perpendicular wavenumbers \\ \hline
    $-64.10\rho_{ti}\leq x \leq 64.10\rho_{ti}$\\
    $-62.83\rho_{ti}\leq y \leq 62.83\rho_{ti}$   \\
    $-\pi \leq z\leq \pi$    \\
    $-4v_{ta}\leq v_\parallel \leq 4v_{ta}$    \\
    $0\leq \mu B_0/T_a \equiv m_av_\perp^2/2T_a\leq 8$    \\ 
$-4.70\rho_{ti}^{-1}\leq k_x\leq4.70\rho_{ti}^{-1}, (k_{x,\mathrm{min}}=0.049\rho_{ti}^{-1})$ \\
$-1.55\rho_{ti}^{-1}\leq k_y\leq1.55\rho_{ti}^{-1}, (k_{y,\mathrm{min}}=0.050\rho_{ti}^{-1})$ \\ \hline
\\ 
\hline
Grid points in $(x, y, z, v_\parallel, \mu)$\\ \hline
$288\times96\times64\times64\times32$ \\
$288\times96\times256\times64\times32$ \\ \hline
\label{tab:resolution}
\end{tabular}
\end{minipage}
\end{table*}
When this model is valid, 
the values of the terms on the right-hand side 
of Eq.~(\ref{eq:EBE_wavenumber}) divided by 
$\langle \langle |\phi_{\bm{k}_\perp}|^2 \rangle \rangle$ 
should not change whether linear or  nonlinear simulations are 
performed to evaluate them. 
On the other hand, when 
divided by $\langle \langle |\phi_{\bm{k}_\perp}|^2 \rangle \rangle$, 
the time-derivative term and the nonlinear entropy transfer term 
on the left-hand side of Eq.~(\ref{eq:EBE_wavenumber}), 
take different values between the linear and nonlinear cases. 
The time-derivative term is dominant and the nonlinear entropy transfer vanishes 
in the linear case, 
while the former is negligible and the latter is dominant in the  nonlinear case. 
Here, the ratio of the collisional dissipation to the squared potential, 
$D_{a\bm{k}_\perp}/\langle \langle |\phi_{\bm{k}_\perp}|^2 \rangle \rangle$, 
is assumed to take the same value. 
Then, it is concluded from the above-mentioned discussion of 
Eq.~(\ref{eq:EBE_wavenumber}) based on the quasilinear model  
that
the ratio of the time-derivative term divided by 
$\langle \langle |\phi_{\bm{k}_\perp}|^2 \rangle \rangle$
in the linear case should be the same as 
the value of  
$\mbox{} - N_{a\bm{k}_\perp} / \langle \langle |\phi_{\bm{k}_\perp}|^2 \rangle \rangle$
in the nonlinear case in order to keep the balance between the left- and right-hand sides of 
Eq.~(\ref{eq:EBE_wavenumber}). 

We here point out the resemblance of Eq.~(\ref{eq:EBE_wavenumber}) to the well-known Landau equation of the weakly nonlinear theory for fluid dynamic systems,\cite{Landau,Drazin}
\begin{equation}
\label{Landau_eq}
\frac{d |A|^2}{dt} = 2 \gamma |A|^2 - \alpha |A|^4
\end{equation}
where $|A|$ and $\gamma$ are the amplitude of the dominant mode and its linear growth rate, respectively, and $\alpha |A|^4$,  with a positive constant $\alpha$, represents the nonlinear effect which causes the saturation of the mode. 
In the linearly growing phase of the mode, the time derivative term on the left-hand side of Eq.~(\ref{Landau_eq})  equals the first term on the right-hand side while in the steady state, the second  nonlinear saturation term on the right-hand side balances with the first term.
This exchange of the roles between the time-derivative and nonlinear terms in Eq.~(\ref{Landau_eq}) is common to the process described about the quasilinear model using Eq.~(\ref{eq:EBE_wavenumber}). 

The summation of Eq.~(\ref{eq:EBE_wavenumber}) over species can be written as
\begin{eqnarray}
& & 
\frac{\partial}{\partial t} \sum_a
\Big\langle\Big\langle 
T_a \int d^3v 
\frac{| f_{a\bm{k}_\perp} |^2}{2f_{Ma}}
+ \frac{e_a}{2} \big( n_{a \bm{k}_\perp}^*  \phi_{\bm{k}_\perp}
+ \frac{1}{c} n_a 
\bm{u}_{a \bm{k}_\perp}^* \cdot A_{\bm{k}_\perp}
\big)             
\Big\rangle\Big\rangle  
\nonumber \\
&&= \frac{\partial}{\partial t}\Big[ \sum_a T_a \Big\langle\Big\langle \int d^3v \frac{ 
\left| f_{a\bm{k}_\perp} \right|^2}{2f_{Ma}}\Big\rangle\Big\rangle + \frac{1}{8\pi} \big\langle\big\langle 
|\bm{E}_{\bm{k}_\perp}|^2+|\bm{B}_{\bm{k}_\perp}|^2\big\rangle\big\rangle\Big]
\nonumber \\
&  & 
= 
 \sum_a \bigg[ \frac{T_a \Gamma^{\rm turb}_{a\bm{k}_\perp}}{L_{pa}}+\frac{q^{\rm turb}_{a\bm{k}_\perp}}{L_{Ta}} 
+  T_a N_{a\bm{k}_\perp} + T_a D_{a\bm{k}_\perp}\bigg]
,
\label{eq:EBE_timederivative}
\end{eqnarray}
where Eqs.~(\ref{eq:Poisson})--(\ref{eq:B_para}) are used. 
We can see that 
terms inside the brackets on the second line of Eq.(\ref{eq:EBE_timederivative}) 
represent the fluctuation entropy  and the electromagnetic energy which 
are all positive,
so that 
their time-derivative terms are positive in the unstable wavenumber region  
in the linear phase of the time evolution of the fluctuation. 
The magnetic fluctuations can be neglected in a low beta plasma, and 
the nonadiabatic part of the electron distribution function is small for ITG 
instability. 
Then, it is considered from Eq.~(\ref{eq:entropy_and_energy}) that,  in the case of the ITG mode, 
the contribution from electrons to the time-derivative part 
in Eq.(\ref{eq:EBE_timederivative}) is negligible and 
ions contribute dominantly. 
Therefore, the time-derivative part on the left-hand side of Eq.~(\ref{eq:EBE_wavenumber}) is 
expected to be positive for ions in the linear phase, 
which implies that  the nonlinear entropy transfer term for ions  in the 
unstable wavenumber region should be negative,  $N_{i \bm{k}_\perp}<0$,  
in the nonlinear steady state
according to the quasilinear argument given earlier. 
On the other hand, 
the summation of $N_{i \bm{k}_\perp}$ over 
the linearly stable wavenumber region
should be positive 
because $\sum_{\bm{k}_\perp} N_{i\bm{k}_\perp}=0$. 
Thus, the entropy due to the fluctuations is transferred from 
the unstable wavenumber region to the stable one.

The discussion of quasilinear weights with entropy balance is expected to apply not only to tokamaks but also to more generic 3D geometries.
In Sec.~\ref{subsec:EB_Quasilinear}, the relative magnitude of each term in the entropy 
balance equation, Eq.~(\ref{eq:h_wavenumber}), is evaluated by the linear and 
nonlinear gyrokinetic simulations, and 
the speculations about the quasilinear model described above 
are examined. 
It should be pointed out that in this work, we discuss only quasilinear weights and do not investigate the saturation rule required to predict fluxes.
%not present a saturation rule which is necessary for developing a quasilinear model. 
%It should be pointed out that only specific plasma case in tokamak geometry is investigated in this work and examinations across parameter space for plasma and geometries are needed to apply the hypothesis to the quasilinear modeling. 
%

\section{Simulation Results}\label{sec:3}
\subsection{Simulation settings}\label{subsec:3A}

\begin{figure*}[tbp]
    \centering
    \includegraphics[keepaspectratio, scale=0.65]{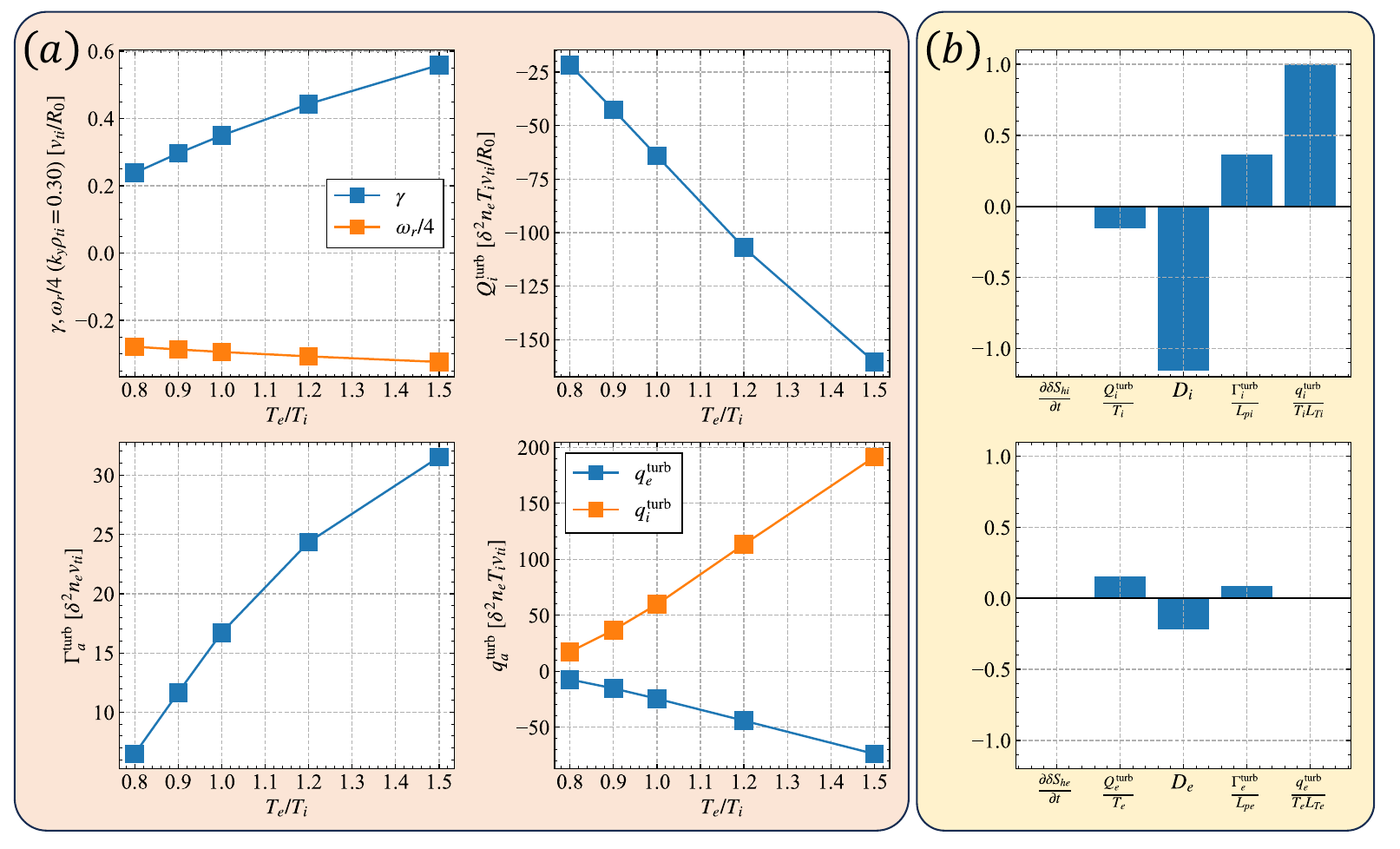}
    \caption{(a) The linear growth rate $\gamma$ and frequency $\omega_r$ at $(k_x\rho_{ti}, k_y\rho_{ti})=(0, 0.30)$, 
the turbulent ion heating $Q^{\rm turb}_i$, 
the turbulent particle flux $\Gamma^{\rm turb}_e =\Gamma^{\rm turb}_i$, 
and the turbulent heat fluxes $q^{\rm turb}_a$ $(a=e,i)$ 
plotted as functions of the temperature ratio $T_e/T_i$. 
(b) Comparison of all terms in the entropy balance equation, Eq.~(\ref{eq:EBequation_h}), in the saturated state of the ITG turbulence for $T_e/T_i=1.0$. 
All terms are normalized by the entropy production $q_i^{\rm turb}/{T_iL_{Ti}}$ due to the turbulent ion heat flux.}
    \label{fig:nonlinearsimulation_tau}
\end{figure*}

In this research, microturbulence simulations are performed  by the GKV code\cite{GKV}, 
which solves the gyrokientic equation for the perturbed distribution function based on the Eulerian scheme in $(k_x, k_y, z, v_\parallel, \mu)$ space.
The nonlinear term is evaluated in the real space, and is transformed back to the wavenumber space by means of 2D Fast Fourier Transform algorithm and the 2/3 rule in $(k_x,k_y)$.
It employs the local flux-tube domain where the background densities, temperatures, and their gradients are fixed. 
While we do not deal with 3D geometries here, microturbulence in herical systems has also been studied by GKV\cite{herical, herical1}.

The flux tube coordinates for a low-$\beta$, large aspect ratio, axisymmetric torus with concentric circular cross sections, $x=r-r_0, y=r_0/q_0\left(q(r)\theta-\zeta\right),$ and $z=\theta$ are used in this work, where $r$, $\theta$, and $\zeta$ are minor radius, poloidal angle, and toroidal angle, respectively. 
The subscript $0$ denotes parameters at the center of flux tube. 
The perpendicular wavenumber is given by $\bm{k}_\perp=\left( k_x +\hat{s} z k_y\right) \bm{e}_r +k_y \bm{e}_\theta$. 
Here, $k_x$ and $k_y$ are wavenumbers in the directions of $\nabla x$ and $\nabla y$, respectively, while $\bm{e}_r$ and $\bm{e}_\theta$ are unit vectors parallel to $\nabla r$ and $\nabla \theta$, respectively.
We here focus on the Ion Temperature Gradient (ITG) mode turbulence in tokamak plasmas so that 
the electron temperature gradient is set to zero, $R_0/L_{Te}=0$, where $R_0$ represent the major radius. 
Plasma and field parameters used in the simulations are shown in Tab.~\ref{tab:Plasma and field parameters}.
Most of them are the same values as in the Cyclone DIII-D base case\cite{Dimits}. 
\begin{comment}
%
Here,  $R_0, \epsilon_r, q_0$, and $\hat{s}=(r/q)(dq/dr)$ 
represent the major radius, inverse aspect ratio, safety factor, and magnetic shear, respectively. 
%
The flux tube coordinates $x=r-r_0, y=r_0/q_0\left(q(r)\theta-\zeta\right),$ and $z=\theta$ are used in a low-$\beta$, large aspect ratio, axisymmetric torus with circular concentric flux surfaces where $r$, $\theta$, and $\zeta$ are minor radius, poloidal angle, and toroidal angle, respectively. 
%
The subscript $0$ denotes parameters at the center of flux tube. 
% 
The perpendicular wavenumber is given by $\bm{k}_\perp=\left( k_x +\hat{s} z k_y\right) \bm{e}_r +k_y \bm{e}_\theta$. 
%
Here, $k_x$ and $k_y$ are wavenumbers in the directions of $\nabla x$ and $\nabla y$, respectively, while $\bm{e}_r$ and $\bm{e}_\theta$ are unit vectors parallel to $\nabla r$ and $\nabla \theta$, respectively.
\end{comment}
%
The ion beta value is set to $\beta_i=1\times 10^{-4}$ for which 
the electrostatic approximation is valid. 
In simulations performed here, $B_{\parallel \bm{k}_\perp}$ is neglected, although $A_{\parallel \bm{k}_\perp}$ is retained in order to avoid numerical difficulty due to very rapid electrostatic waves called the $\omega_H$ mode \cite{Lee}. 
The Lenard-Bernstein collision operator\cite{Lenard} is used here because it takes less computation time than more rigorous collision models.
However, we expect that the collision model does not influence results from the present study where the normalized collision frequency is set to $\nu^*_{ii}\equiv R_0 q \nu_{ii}/(\epsilon^{3/2}v_{ti}) = 0.068$, which is much smaller than the growth rates of the ITG modes in the present study. 
%[change]
Since collisional energy exchange is proportional to the temperature difference between electrons and ions, the temperature ratio is set in a range from $T_e/T_i =0.80$ to $1.5$ in order to compare turbulent and collisional energy exchanges.
%

\begin{figure*}[t]
    \centering
    \includegraphics[keepaspectratio, scale=0.51]{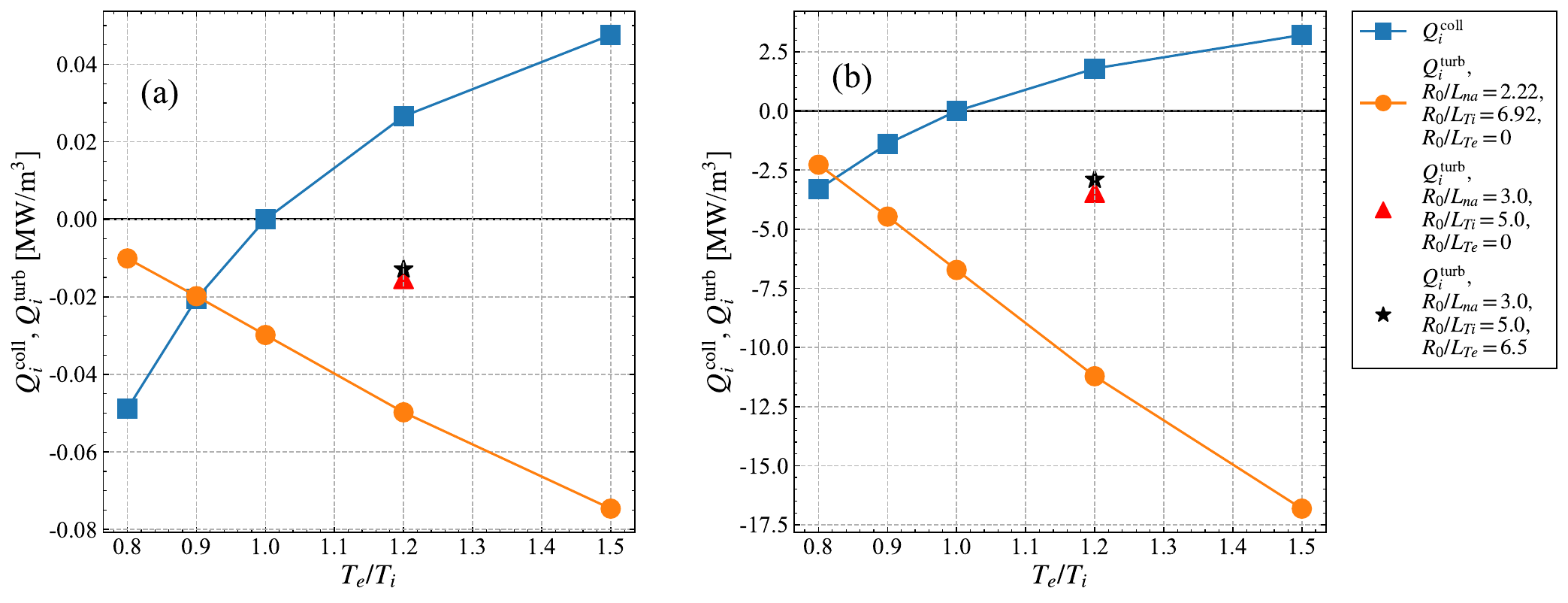}
    \caption{Comparison of energy exchanges due to Coulomb collisions (blue squares) and turbulence (orange circles). 
The left figure (a) shows the case of $\delta\equiv\rho_{ti}/R_0=9.6\times 10^{-4}$, $n_e=2.0\times 10^{19} [\rm{m^{-3}}]$, $T_i=0.9[\rm{keV}]$, and the right one (b) shows the case of $\delta=1.7\times10^{-3}$, $n_e=2.25\times 10^{20} [\rm{m^{-3}}]$, $T_i=3.0[\rm{keV}]$. 
Red triangles and black stars indicate turbulent energy exchanges obtained from simulations for $T_e/T_i = 1.2$ and $R_0/L_{na}=3.0$ using different temperature gradients given by 
$(R_0/L_{Ti}, R_0/L_{Te} ) = (5.0, 0)$ and  $(R_0/L_{Ti}, R_0/L_{Te} ) = (5.0, 6.5)$, respectively.}
    \label{fig:energyexchangecomparison}
\end{figure*}

\begin{comment}
%
For the nonlinear simulations, the domain sizes are basically set to $-64.10\rho_{ti}\leq x \leq 64.10\rho_{ti}, -62.83\rho_{ti}\leq y \leq 62.83\rho_{ti},-\pi \leq z\leq \pi, -4v_{ta}\leq v_\parallel \leq 4v_{ta}$, and $0\leq \mu B_0/T_a \equiv m_av_\perp^2/2T_a\leq 8$. 
%
Numbers of grid points in $(x, y, z, v_\parallel, \mu)$  are $288\times96\times64\times64\times32$ or $288\times96\times256\times64\times32$.
%
The resolved perpendicular wavenumbers are $-4.70\rho_{ti}^{-1}\leq k_x\leq4.70\rho_{ti}^{-1}$ and $-1.55\rho_{ti}^{-1}\leq k_y\leq1.55\rho_{ti}^{-1}$ with the minimum wavenumbers $(k_{x,\mathrm{min}}, k_{y,\mathrm{min}})=(0.049\rho_{ti}^{-1}, 0.050\rho_{ti}^{-1})$.
%
\end{comment}
%
The resolution settings are shown in Tab.~\ref{tab:resolution}.
High resolution for the parallel coordinate $z$ is used in Fig.~\ref{fig:nonlinearsimulation_tau} (b) and Fig.~\ref{fig:quasilinear} to suppress an error of entropy balances by a hyper diffusion term\cite{Hyper Diffusion}.
There is no significant difference in turbulent fluxes and energy exchange when using either low or high resolutions for $z$.
The output data shown in this paper are normalized by following Ref.~\cite{GKV} except for Fig.~\ref{fig:energyexchangecomparison}.

\subsection{Heat flux, turbulent energy exchange, and entropy balance}\label{subsec:3B}

Here we perform linear and nonlinear simulations where the temperature ratio is varied as a parameter. 
Fig.~\ref{fig:nonlinearsimulation_tau}(a) shows the linear growth rate for $(k_x\rho_{ti}, k_y\rho_{ti})=(0.00, 0.30)$, particle and heat fluxes, and energy exchange as functions of the temperature ratio $T_e/T_i$. 
The results except for the linear growth rate are obtained by taking a time average in a steady state of turbulence. 
It is seen that, as $T_e/T_i$ increases,  the linear growth rate, the absolute values of particle and heat fluxes and turbulent energy exchange increase. 

The ratio of each entropy balance term to the entropy production term caused by the ion heat flux $q^{\rm turb}_i/T_iL_{Ti}$ is shown in Fig. \ref{fig:nonlinearsimulation_tau}(b). 
The numerical error in the entropy balance defined by the difference between the left- and right-hand sides of Eq.~(\ref{eq:EBequation_h}) is within 6\% of $q^{\rm turb}_i/T_iL_{Ti}$ for ions and 3\% for electrons. 
In the case of ions, the particle and heat fluxes generate entropy while the turbulent energy exchange and collisions reduce entropy, thus maintaining balance. 
For electrons, on the other hand, the electron heat flux does not appear as an entropy-producing term because of no electron temperature gradient, although the turbulent energy exchange plays a major role in the entropy production in addition to the electron particle flux. 
The generated entropy is dissipated by collisions, 
and the entropy of turbulent fluctuations in the electron distribution function is kept in 
a steady state. 
The same result as described above is also reported in ref. \cite{Maeyama}. 
 
In ITG turbulence, 
the turbulence entropy of ions is 
generated primarily by the product of the ion heat flux and the ion temperature gradient 
and secondly by that of the particle flux and the ion pressure gradient. 
The ion turbulence entropy is lost mainly through collisions, although it is partially transferred to the electron turbulence entropy by the turbulent energy exchange. 
Electrons increase their turbulence entropy by the energy transfer from ions and 
the particle flux, while they reduce it through collisions. 
Thus, the total turbulence entropy balance in the steady state of the ITG turbulence 
is maintained by the turbulent energy transfer from ions to electrons, which carries 
the excess of the stronger ion entropy production to the weaker electron portion.

\subsection{Comparison between collisional and turbulent energy exchanges}\label{subsec:3C}
\begin{figure*}[tbp]
    %\centering
    \includegraphics[keepaspectratio, scale=0.55]{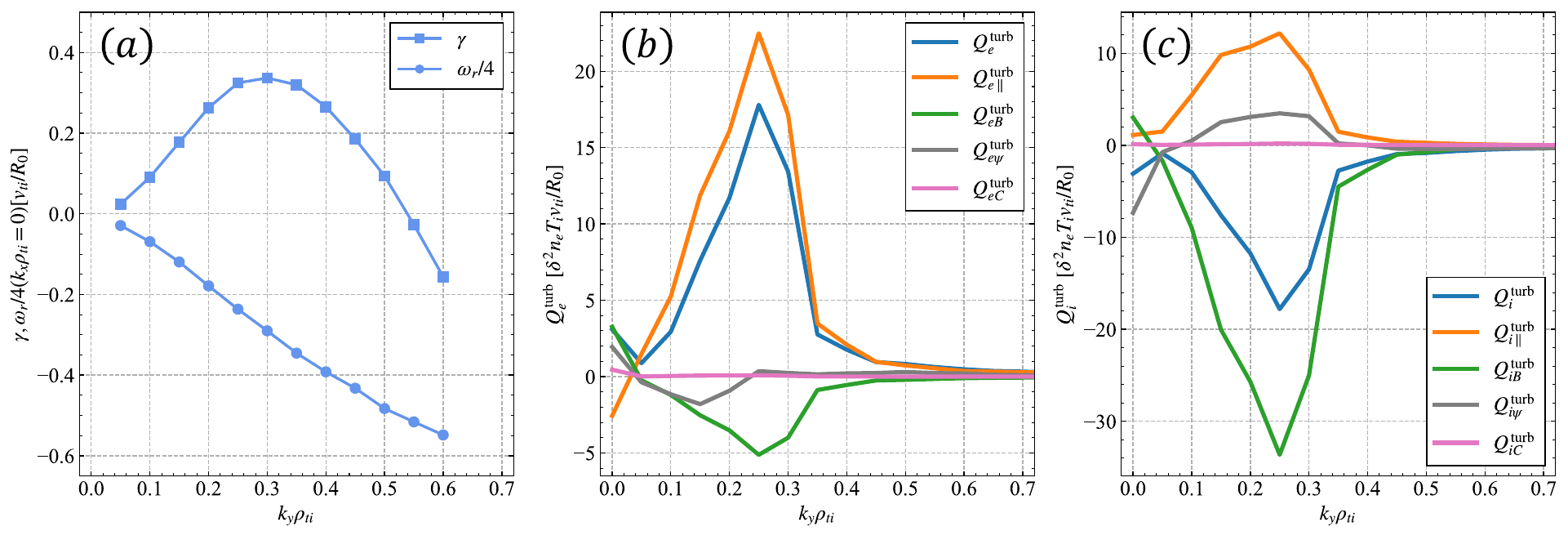}
    \label{fig:componentspectrum}
    \\
    %\centering
    \includegraphics[keepaspectratio, scale=0.55]{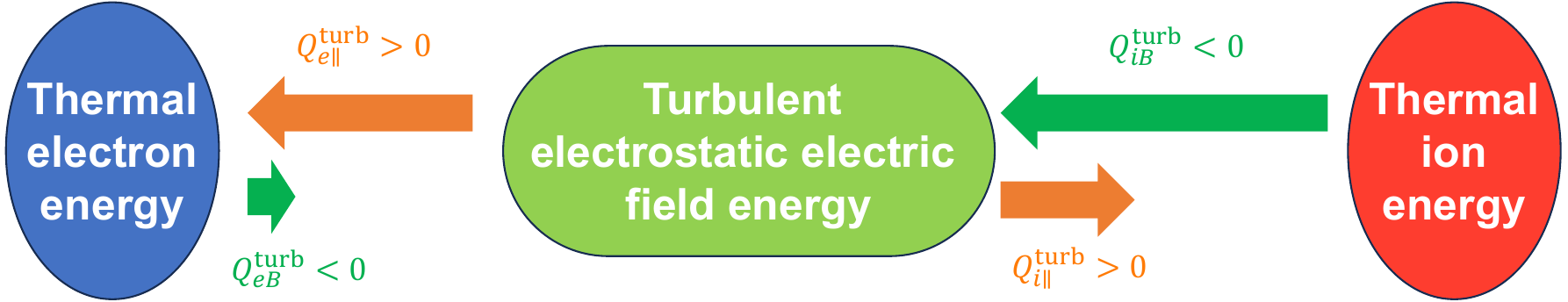}
    \label{fig:componentspectrum_picture}
    \caption{The wavenumber spectra of the linear growth rate and frequency (a) and the turbulent energy transfer terms in Eqs.~(\ref{eq:TurbulentEnergyExchange_PowerFlow})--(\ref{eq:collision_psi}) for electrons (b) and ions (c) in the case of $T_e/T_i=1.0$. 
The spectra are given as functions of $k_y \rho_{ti}$ obtained by summing over $k_x$. 
The peaks and valleys of the turbulent energy transfer terms are found at a wavenumber lower than that of the linearly most unstable mode. 
The directions and magnitudes of the turbulent energy transfer terms are represented by arrows in the bottom figure which schematically shows the role of turbulent energy transfer terms based on Eqs.~(\ref{eq:transport equation}), (\ref{eq:TurbulentEnergyExchange_PowerFlow}), and (\ref{psidth}) with magnetic fluctuations neglected. The electron heating due to the parallel field denoted by $Q_{e\parallel}^{\rm turb}>0$ and the ion cooling due to the $\nabla B$-curvature drift denoted by  $Q_{iB}^{\rm turb}<0$ are dominant mechanisms in the turbulent energy exchange between electrons and ions in ITG turbulence. 
}
    \label{fig:componentspectrum}
\end{figure*}

%the GKV code normalizes the turbulent energy exchange in gyroboom units.

Here, the results of the turbulent energy exchange shown in Fig.\ref{fig:nonlinearsimulation_tau}(a) are used for comparison with the collisional energy exchange calculated by Eq.~(\ref{eq:Collisional energy exchange}).  
If we treat the Coulomb logarithm as a constant ($\ln \Lambda= 15.5$) and vary the density and temperature with keeping $n/T^2$ fixed, the normalized collision frequency used as an input parameter 
of the GKV code does not change. 
Therefore, 
the results in Fig.~\ref{fig:nonlinearsimulation_tau}(a)
can be used for the two density and temperature conditions with the same value of $n/T^2$
shown in Figs.~\ref{fig:energyexchangecomparison} (a) and (b). 
The results for the density-temperature condition at $r=a/2$ at the DIII-D128913 shot ($0.9[\rm{keV}]$, $2.0\times 10^{19}[\rm{m^{-3}}]$) and for the $3.0[\rm{keV}]$, $2.25\times10^{20} [\rm{m^{3}}]$ are plotted in Figs.~\ref{fig:energyexchangecomparison} (a) and (b), respectively. 
The collisional and turbulent energy transfers from electrons to ions, $Q_i^{\rm coll}$ and $Q_i^{\rm turb}$,  are shown as functions of the temperature ratio $T_e/T_i$ in Fig.\ref{fig:energyexchangecomparison}, where 
we can identify a difference between the directions of collisional and turbulent energy transfers. 
In Coulomb collisions, energy is transferred always from higher temperature particles to lower temperature ones. 
Thus, when $T_e/T_i < 1 (> 1)$ is less (more) than unity, $Q_i^{\rm coll}$ is negative (positive). 
On the other hand, the ITG turbulence always transfers energy from ions to electrons regardless of the value of $T_e/T_i$. 
We can see that $Q_i^{\rm turb} < 0$ even in the equithermal condition $T_e/T_i = 1$ where $Q_i^{\rm coll}=0$ and that $Q_i^{\rm coll}$ and $Q_i^{\rm turb}$ take opposite signs to each other when $T_e/T_i > 1$. 

It can also be confirmed by comparing with Figs.~\ref{fig:energyexchangecomparison} (a) and (b) that as the temperatures for electrons and ions are increased with their ratio $T_e/T_i$ fixed, the energy exchange by turbulence becomes more dominant than that by collision.
Note that, based on the gyrokinetic ordering, 
the turbulent ion heating is  written as 
$
Q_i^{\rm turb} = (\rho_{ti}/R_0)^2 (v_{ti}/R_0) (n T_i ) 
Q^* 
$
where 
$Q^* $ is a dimensionless function of dimensionless variables 
$
T_e/T_i,  R_0/L_n, R_0/L_{Ti}, R_0/L_{Te}, q_0, \hat{s}, \beta, R_0 \nu_i/v_{ti}, \cdots 
$
which are used as input parameters of the local flux-tube simulation. 
This expression is combined with Eq.~(\ref{eq:Collisional energy exchange}) to obtain
\begin{eqnarray}
\frac{Q_i^{\rm turb}}{Q_i^{\rm coll}}
& = & 
\frac{m_i/m_e}{ 3(T_e/T_i - 1 )} \frac{v_{ti}}{R_0 \nu_e} 
\left(\frac{\rho_{ti}}{R_0}\right)^2 
\nonumber \\ 
& & \mbox{} 
\times 
Q^* \left(
\frac{T_e}{T_i},  \frac{R_0}{L_n}, \frac{R_0}{L_{Ti}}, \frac{R_0}{L_{Te}}, q_0, \hat{s},  \beta, 
R_0 \nu_i/v_{ti}, \cdots 
\right)
\nonumber \\ 
& \propto & \frac{T_i}{B^2 R_0^3 (R_0 \nu_i/v_{ti})} \frac{Q^*}{ (T_e/T_i - 1 )}
\nonumber \\ 
& \propto & \frac{T_i^3}{n B^2 R_0^3} \frac{Q^*}{ (T_e/T_i - 1 )}
.
\end{eqnarray}
When the normalized values 
$
T_e/T_i,  R_0/L_n, R_0/L_{Ti}, R_0/L_{Te},q_0,$ $\hat{s}, \beta, R_0 \nu_i/v_{ti}, \cdots
$
are regarded as constants, 
$Q^*$ is also a constant and 
the ratio $Q_i^{\rm turb}/Q_i^{\rm coll}$ between the turbulent and collisional energy exchanges is proportional to the temperature. 
This temperature dependence of $Q_i^{\rm turb}/Q_i^{\rm coll}$ is seen by comparing the results 
in Figs.~\ref{fig:energyexchangecomparison} (a) and (b). 
If taking account of the temperature dependence of $R_0 \nu_i/v_{ti}$ and assuming $Q^*$ 
to depends weakly on $R_0 \nu_i/v_{ti}$, 
$Q_i^{\rm turb}/Q_i^{\rm coll}$ is proportional to the cubic of the temperature. 
In the case of high plasma temperatures with $T_e/T_i > 1$, 
the net energy transfer from lower-temperature ions to higher-temperature electrons 
can occur, contrary to conventional thought.

It is reported in Ref.\cite{Candy} that the turbulent energy exchange has a negligible effect on the simulation for predicting the temperature profile in the case of DIII-D128913. 
Here, we compare that case with the Cyclone D-III D base case (CBC) used in our simulations. 
The normalized density and temperature gradients in DIII-D128913 are estimated as $R_0/L_{na}=3.0$,  $R_0/L_{Ti}=5.0$, which is smaller than  $R_0/L_{Ti}=6.92$ at $r/a = 0.5 $ in CBC, and $R_0/L_{Te}=6.5$ at the minor radius $r/a = 0.5 $ from Ref.\cite{Candy, White}.
In Fig.~\ref{fig:energyexchangecomparison}, turbulent energy exchanges obtained from simulations for $T_e/T_i = 1.2$ and $R_0/L_{na}=3.0$ using $(R_0/L_{Ti}, R_0/L_{Te} )  = (5.0, 0)$ and  $(R_0/L_{Ti}, R_0/L_{Te} ) = (5.0, 6.5)$ are plotted by red triangles and black stars, respectively.
Interestingly, these plots indicate that the dependence of the turbulent energy exchange on $R_0/L_{Te}$ is weak while the turbulent transport fluxes are found to increase significantly with increasing $R_0/L_{Te}$. 
It is speculated that, even though both ion and electron turbulence entropy production due to transport fluxes increase, their difference stays the same so that the entropy balance for each species is maintained with the turbulent energy exchange unaltered.
The black star in the left figure for $n_e=2.0\times 10^{19} [\rm{m^{-3}}]$ and $T_i=0.9[\rm{keV}]$
corresponds to the conditions at $r/a = 0.5$ in DIII-D128913, which shows that the magnitude of the turbulent energy exchange becomes significantly smaller that that of the collisional energy exchange. 
Thus, in this case, the turbulent energy exchange has only a small influence, which is consistent with the result in Ref.\cite{Candy}. 

On the other hand, if the ion temperature gradient is so large that the turbulence has a dominant effect on the energy exchange, a net energy transfer can occur from ions to electrons even for $T_e > T_i$. 
In particular, since the energy exchange due to the ITG turbulence acts to prevent the energy transfer from electrons to ions, 
the ITG turbulence is undesirable for fusion reactors in that it interferes with the ion heating by the alpha-heated electrons, as well as degrading the ion energy confinement through enhancing the energy transport toward the outside of the device. 

\subsection{Spectral analysis of the turbulent energy exchange}\label{subsec:3D}

\begin{figure*}[tbp]
    \centering
    \includegraphics[keepaspectratio, scale=0.6]{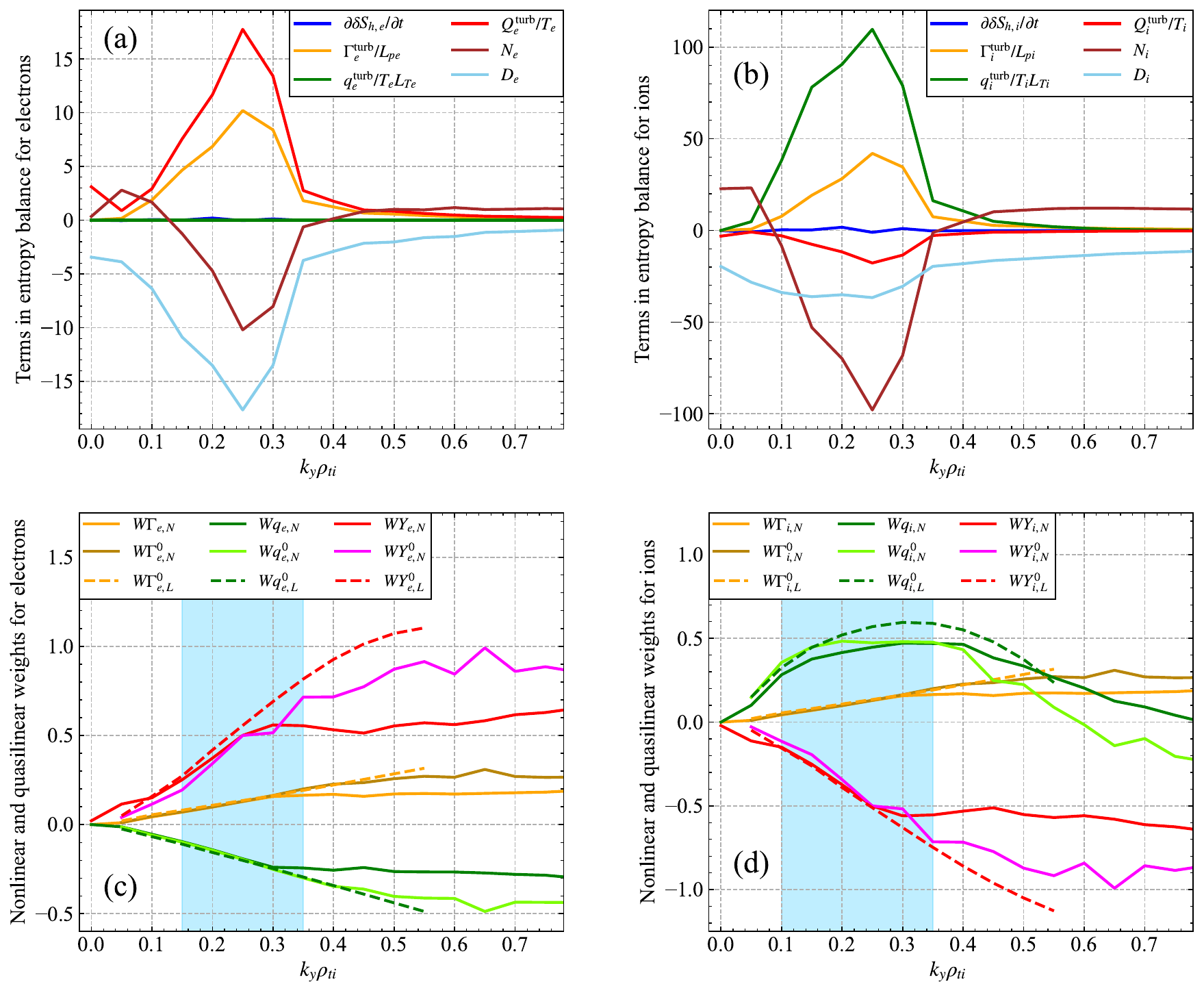}
    \caption{The wavenumber spectra of terms in the entropy balance equation in Eq.~(42) for electrons (a) and ions (b) in the case of $T_e/T_i = 1.0$. The spectra are given as functions of $k_y \rho_{ti}$ obtained by summing over $k_x$. They are evaluated in the steady state of turbulence obtained by nonlinear simulation. The ratios of the turbulent particle and heat transport fluxes and the turbulent energy exchange to the squared amplitude of the electrostatic potential obtained by linear and nonlinear simulations are shown for electrons (c) and ions (d).  The nonlinear entropy transfer terms  $N_{ek_y}$ and $N_{ik_y}$ are negative in the wavenumber regions colored in sky blue. Dashed lines in (c) and (d) represent the ratios obtained by linearly unstable modes with $k_x  = 0$.}
    \label{fig:quasilinear}
  \end{figure*}

Next, we examine each component of the wavenumber spectrum of the turbulent energy exchange shown in Eq.(\ref{eq:TEE_parts}). 
Figure~\ref{fig:componentspectrum} (a) shows the wavenumber spectra of the linear growth rate and frequency.
The turbulent energy transfer terms in Eqs.~(\ref{eq:TurbulentEnergyExchange_PowerFlow})--(\ref{eq:collision_psi}) in the case of $T_e/T_i = 1.0$ are shown for electrons and ions in Figs.~\ref{fig:componentspectrum} (b) and (c), respectively. 
First, the collisional components $Q^{\mathrm{turb}}_{eC\bm{k}_\perp}$ and $Q^{\mathrm{turb}}_{iC\bm{k}_\perp}$ have little effect on the turbulent energy exchange. 
In addition, both electrons and ions show positive values of the parallel-heating components 
$Q^{\mathrm{turb}}_{a\parallel \bm{k}_\perp}>0$ $(a=e,i)$ , and negative values of the components 
$Q^{\mathrm{turb}}_{aB \bm{k}_\perp}<0$  $(a=e,i)$ caused by the product of the perpendicular field and the $\nabla B$ -curvature drift velocity in the wavenumber region where the ITG mode is linearly unstable. 
At the bottom of Fig.\ref{fig:componentspectrum}, the roles of turbulent energy transfer terms in increasing or decreasing the thermal electron and ion energies, and the turbulent electrostatic electric field energy are schematically shown, based on Eqs.~(\ref{eq:transport equation}), (\ref{eq:TurbulentEnergyExchange_PowerFlow}), and (\ref{psidth}) with the magnetic fluctuations neglected. 
We can see that the perpendicular ion cooling represented by $Q^{\mathrm{turb}}_{iB} < 0$ is dominant and overcomes the parallel ion heating $Q^{\mathrm{turb}}_{i\parallel} > 0$, which leads to the net turbulent ion cooling shown by $Q_i^{\rm turb} < 0$. 
On the other hand, the parallel heating $Q^{\mathrm{turb}}_{e\parallel}>0$ is dominant for electrons. 
Thus, the perpendicular ion cooling and the parallel electron heating are found to be the main mechanisms in the turbulent energy transfer from ions to electrons in a steady state of ITG turbulence. 

This perpendicular ion cooling is connected with the mechanism of ITG instability. 
In low beta plasmas, the $\nabla B$-curvature drift can be expressed as 
 $\bm{v}_{da} =\bm{b}\times\left(v_\parallel^2+\mu B/m_a\right)\nabla B/\Omega_a B$. 
Then, 
under the electrostatic approximation,  $Q^{\mathrm{turb}}_{iB \bm{k}_\perp}$ in Eq.~(\ref{eq:perp_heating})  is rewritten as 
\begin{equation}
    Q^{\mathrm{turb}}_{iB \bm{k}_\perp} = 
\Re\Bigg\langle\Bigg\langle 2  P_{i\bm{k}_\perp}^* 
\left(
 \frac{c}{B} \bm{E}_{\bm{k}_\perp}\times \bm{b}\right)\cdot 
\nabla \ln{B}
\Bigg\rangle\Bigg\rangle
\label{eq:perp_heating_ITG}
\end{equation}
where $\bm{E}_{\bm{k}_\perp} \equiv -i \bm{k}_{\perp}  \hat{\phi}$ is the electric field and 
$P_{i \bm{k}_\perp} \equiv \int d^3v\left( m_i v_\parallel^2+\mu B\right) f_{i\bm{k}_\perp}/2 $ roughly represents ion pressure perturbation.  
The ITG mode is destabilized at the outside of the torus (or the bad curvature region) where 
the ion pressure perturbation is amplified by the outward ${\bf E} \times {\bf B}$ flow from the inner hot plasma 
region. 
Therefore, the phases of $P_{i \bm{k}_\perp}$ and $(c\bm{E}_{\bm{k}_\perp}\times \bm{b}/B )\cdot 
\nabla \ln{B}$ become opposite to each other, and accordingly, $Q^{\mathrm{turb}}_{iB}<0$ is expected for the ITG instability from Eq.~(\ref{eq:perp_heating_ITG}). 
We also see from Eq.~(\ref{eq:perp_heating_ITG}) that $- Q^{\mathrm{turb}}_{iB}$ represents the outward energy flow in the bad curvature. 
Thus,  $Q^{\mathrm{turb}}_{iB}<0$ means the outward energy transport $- Q^{\mathrm{turb}}_{iB} > 0$ due to the ITG turbulence. 

In Fig.~\ref{fig:componentspectrum}, the nonlinear interaction between different wavenumbers is shown by the components $Q^{\mathrm{turb}}_{e\psi \bm{k}_\perp}$ and $Q^{\mathrm{turb}}_{i\psi \bm{k}_\perp}$. 
They show that the energy in the unstable wavenumber region is carried to the zonal flow mode with 
$k_y = 0$. 
This implies that the zonal flows play a significant role in the nonlinear saturation of the ITG turbulence. 
It is also confirmed that $\sum_{\bm{k}_\perp} Q^{\mathrm{turb}}_{a\psi \bm{k}_\perp} = 0$ 
$(a=e,i)$ and the nonlinear interaction causes no net energy production.

\subsection{Correlation between results from linear and nonlinear simulations}
\label{subsec:EB_Quasilinear}

In Secs.~III B and C, we investigated the characteristics and the physical mechanism of the  energy exchange due to the ITG turbulence. 
In particular, the comparison with the collisional energy exchange has clarified that the turbulent energy exchange can play a dominant role in  the energy exchange between electrons and ions in low collision or high temperature plasmas. 
Therefore, it is necessary to take account of the effects of the  turbulent energy exchange along with those of the particle and heat fluxes for reliable predictions of the global density and temperature profiles in future fusion reactors. 
In this subsection, the nonlinear simulation results of the ratio of the turbulent energy exchange to the squared amplitude of the electrostatic potential, are compared with the linear simulation results in order to examine the validity of the quasilinear model for predicting turbulent energy exchange.
Equation~(\ref{eq:Candy}) is used here for evaluating the turbulent energy exchange from the linear simulations. 

Figures.~\ref{fig:quasilinear} (a) and (b) show the wavenumber spectra of all terms in the entropy balance equations for electrons and ions, respectively. 
They are evaluated in the steady state of turbulence obtained by the nonlinear simulation for $T_e/T_i=1.0$.
The turbulent ion heat flux under the ion temperature gradient makes the largest contribution to the entropy production in the unstable wavenumber region 
where the particle flux under the pressure gradient also produces 
the entropy for both electrons and ions. 
On the other hand, no contribution is made by the electron heat flux because the electron temperature gradient is set to zero in the present simulation condition. 
The entropy produced by the unstable modes in the ITG turbulence is transferred to the zonal flow modes around $k_y = 0$  and to the high-wavenumber modes, while the collisional entropy dissipation represented by $D_e < 0$ and $D_i <0$ occurs in a wide wavenumber range, which 
maintains the detailed entropy balance in each wavenumber. 

The ratios of the turbulent particle and heat transport fluxes and the turbulent energy exchange to the squared amplitude of the electrostatic potential are plotted for electrons and ions in Figs~\ref{fig:quasilinear} (c) and (d), respectively. 
Here, it should be recalled that these ratios obtained by linear simulations are called quasilinear weights. 
On the other hand, those obtained by nonlinear simulations are called nonlinear weights here. 
Dashed and solid curves represent the quasilinear and nonlinear weights, respectively.
Solid curves labeled $\left( W\Gamma_{a,N}, Wq_{a,N}, WY_{a,N} \right)$ in Figs.~\ref{fig:quasilinear} (c) and (d) show the nonlinear weights as functions of $k_y \rho_{ti}$ calculated by
\begin{eqnarray}
W\Gamma_{a,N}(k_y)&=&\frac{T_i^2}{e^2n_ev_{ti}}\frac{\sum_{k_x}\langle \langle \tilde{\Gamma}_{a\bm{k}_\perp,N}\rangle\rangle}{\sum_{k_x}\langle \langle |\phi_{\bm{k}_\perp, N}|^2 \rangle\rangle}, \label{eq:WG_N}\\
Wq_{a,N}(k_y)&=&\frac{T_i}{e^2n_ev_{ti}}\frac{\sum_{k_x}\langle \langle \tilde{q}_{a\bm{k}_\perp,N}\rangle\rangle}{\sum_{k_x}\langle \langle |\phi_{\bm{k}_\perp, N}|^2 \rangle\rangle}, \label{eq:Wq_N}\\
WY_{a,N}(k_y)&=&\frac{T_iR_0}{e^2n_ev_{ti}}\frac{\sum_{k_x}\langle \langle \tilde{Y}_{a\bm{k}_\perp,N}\rangle\rangle}{\sum_{k_x}\langle \langle |\phi_{\bm{k}_\perp, N}|^2 \rangle\rangle}, \label{eq:WY_N}
\end{eqnarray}
where $\tilde{\Gamma}_{a\bm{k}_\perp}$, $\tilde{q}_{a\bm{k}_\perp}$, and $\tilde{Y}_{a\bm{k}_\perp}$ are defined by the real parts of integrals inside the double average over the ensemble and the flux surface $\langle \langle \cdots \rangle \rangle$ of Eqs.~(\ref{eq:particle and heat fluxes}) and (\ref{eq:Candy}), %
\begin{eqnarray}
&&\tilde{\Gamma}_{a\bm{k}_\perp} \equiv \Re\left[  \int d^3v
h_{a\bm{k}_\perp}^*\left(-i\frac{c}{B}\psi_{a\bm{k}_\perp}\bm{k}_\perp\times\bm{b}\right)\cdot\nabla r \right],\\
&&\tilde{q}_{a\bm{k}_\perp} \equiv \Re\left[  \int d^3v
\left(w-\frac{5T_a}{2}\right) h_{a\bm{k}_\perp}^*\left(-i\frac{c}{B}\psi_{a\bm{k}_\perp}\bm{k}_\perp\times\bm{b}\right)\cdot\nabla r \right],\nonumber \\
\\
&&\tilde{Y}_{a\bm{k}_\perp}\equiv \Re\left[\frac{e_a}{2} \int d^3v 
\left(h_{a\bm{k}_\perp}^*\frac{\partial\psi_{a\bm{k}_\perp}}{\partial t} 
- \frac{\partial h^*_{a\bm{k}_\perp}}{\partial t} \psi_{a\bm{k}_\perp}\right)\right]. \label{eq:tildeY}
\end{eqnarray}
On the right-hand side of Eqs.~(\ref{eq:WG_N})--(\ref{eq:WY_N}), coefficients for normalization of each weights are included and the time average in the steady state of turbulence is used instead of the ensemble average to evaluate $\langle \langle \cdots \rangle \rangle$. 
Other curves labeled $\left( W\Gamma^0_{a,N}, Wq^0_{a,N}, WY^0_{a,N} \right)$ and  $\left( W\Gamma^0_{a,L}, Wq^0_{a,L}, WY^0_{a,L} \right)$ represent the nonlinear and quasilinear weights obtained by

\begin{eqnarray}
W\Gamma^0_{a,N}(k_y)&=&\frac{T_i^2}{e^2n_ev_{ti}}\frac{\langle\langle \tilde{\Gamma}_{ak_x=0, k_y, N}\rangle\rangle}{\langle\langle |\phi_{k_x=0,  k_y, N}|^2 \rangle\rangle}, \label{eq:WG0N}\\
Wq^0_{a,N}(k_y)&=&\frac{T_i}{e^2n_ev_{ti}}\frac{\langle\langle \tilde{q}_{ak_x=0, k_y, N} \rangle\rangle}{\langle\langle |\phi_{k_x=0,  k_y, N}|^2 \rangle\rangle}, \label{eq:Wq0N}\\
WY^0_{a,N}(k_y)&=&\frac{T_iR_0}{e^2n_ev_{ti}}\frac{\langle\langle \tilde{Y}_{ak_x=0, k_y, N}\rangle\rangle}{\langle\langle |\phi_{k_x=0, k_y, N}|^2 \rangle\rangle},\label{eq:WY0N}
\end{eqnarray}
and,
\begin{eqnarray}
W\Gamma^0_{a,L}(k_y)&=&\frac{T_i^2}{e^2n_ev_{ti}}\frac{\langle \tilde{\Gamma}_{ak_x=0, k_y, L}\rangle}{\langle |\phi_{k_x=0,  k_y, L}|^2 \rangle}, \label{eq:WG0L}\\
Wq^0_{a,L}(k_y)&=&\frac{T_i}{e^2n_ev_{ti}} \frac{\langle \tilde{q}_{ak_x=0, k_y, L}\rangle}{\langle |\phi_{k_x=0, k_y, L}|^2 \rangle}, \label{eq:Wq0L}\\
WY^0_{a,L}(k_y)&=&\frac{T_iR_0}{e^2n_ev_{ti}}\frac{\langle \tilde{Y}_{ak_x=0, k_y, L}\rangle}{\langle |\phi_{k_x=0,  k_y, L}|^2 \rangle},\label{eq:WY0L}
\end{eqnarray}
where only the $k_x=0$ modes are kept instead of summing over $k_x$.
Here, the subscripts $L$ and $N$ denote the results from the linear and nonlinear simulations, respectively.
On the right-hand side of Eqs.~(\ref{eq:WG0L})--(\ref{eq:WY0L}),  
$\langle \cdots \rangle$ denotes the surface average and the ratios of 
$\langle\tilde{\Gamma}_{ak_x=0, k_y, L} \rangle$, $\langle \tilde{q}_{ak_x=0, k_y, L} \rangle$ and $\langle \tilde{Y}_{ak_x=0, k_y, L} \rangle$ to $\langle \phi_{k_x=0, k_y, L}|^2 \rangle$ are evaluated for the linear unstable mode with the wavenumbers $k_x = 0$ and $k_y$. 

The quasilinear weights for the $k_x=0$ modes show a good agreement to those of the nonlinear weights for  $k_x=0$ in the linearly unstable wavenumber region $0.05 \leq k_y \rho_{ti} \leq 0.5$ [see Fig.~\ref{fig:componentspectrum} (a)]. 
In the areas colored in sky blue, we have $N_{e k_y} < 0$ and $N_{i k_y} < 0$, which indicate that the entropy of the fluctuation at $k_y$ is transferred to other wavenumber regions through nonlinear interaction. 
Both of the nonlinear weights obtained by keeping the only $k_x=0$ modes and by summing over $k_x$ agree well with each other in the colored regions. 
Thus,  in these regions, the nonlinear weights including all $k_x$'s are well approximated by the quasilinear weights for $k_x=0$ within an error margin of 30$\%$ or less.
We find that more than 80$\%$ of the total values of the transport fluxes and the energy exchange over the whole wavenumber space can be accounted for by contributions from the colored wavenumber regions. 
Therefore, nonlinear simulation results of the transport fluxes and the energy exchange can be effectively predicted from the quasilinear weights for $k_x = 0$ multiplied by the squared potential amplitude. 
The model for predicting the magnitude of the potential amplitude is not treated in the present work, although there are many analytical and numerical studies to parameterize the saturation amplitude.\cite{Casati, Bourdelle, Staebler, Toda, Parker} 
The results shown above indicate that it is possible to construct a quasilinear model which can accurately predict both of the turbulent transport fluxes and the turbulent energy exchange. 

\section{Conclusions and Discussion}\label{sec:4}
In this study, the effect of ITG turbulence on the energy exchange between electrons and ions in tokamak plasmas is investigated. 
The ITG  turbulence is found to be dominant in the energy exchange in equithermal or high-temperature plasmas in which collisional energy exchange is negligibly small. 
It is also shown that the direction of net energy transfer can be opposite to that of the collisional one from hotter to colder particle species, since ITG turbulence transfers energy from ions to electrons, even when ions are colder than electrons. 
This result does not contradict with the second law of thermodynamics because the entropy balance is still maintained by the entropy production, mainly due to the ion heat transport from hot to cold regions. 
Therefore, the ITG turbulence is anticipated to prevent energy transfer from alpha-heated electrons to ions, which is considered a primary ion heating mechanism in future reactors. 
The wavenumber spectral analysis reveals that the main physical mechanisms of turbulent energy exchange are the cooling of ions in the $\nabla B$-curvature drift motion and the heating of electrons streaming along the field line, which are caused by the perpendicular and parallel components of the turbulent electric field, respectively.  
In particular, the perpendicular cooling of ions is closely linked to the physical mechanism of ITG instability which drives the ion heat flux.

Since the effect of turbulence on the energy exchange between electrons and ions can possibly overcome that of Coulomb collisions, the turbulent energy exchange as well as the turbulent transport fluxes of particles and heat should be taken into account for predicting global profiles of the density and temperature in future fusion reactors.
In order to examine the predictability of the quasilinear model of the energy exchange and transport fluxes, the quasilinear weights of these turbulent quantities normalized by the squared potential are estimated as functions of the wavenumber by linear simulations. 
It is found that the quasilinear weights of both energy exchange and fluxes agree with the nonlinear simulation results within an error margin of $30\%$ in the wavenumber region, where more than $80\%$ of total energy exchange and fluxes are covered. 
This indicates that we can construct the quasilinear model which is valid for predicting the energy exchange as well as the transport fluxes. 
Although this study has not investigated the saturation model, it would not be difficult to incorporate turbulent energy exchange in existing codes that predict fluxes with a quasilinear model\cite{Staebler, Parker}.
 
The entropy balance in the linearly-growing and nonlinearly-saturated phases of the ITG modes is  examined to understand the conditions for the quasilinear weights estimated from the linear analysis to be applicable to describing the steady state of turbulence.  
The analogy of the entropy balance equation to the Landau equation for weakly nonlinear fluid dynamic systems is noted in that the time-derivative term in the linear phase should be replaced by the nonlinear term in the steady state for the quasilinear weights to keep the same ratios in both phases. 
Then we speculate that the nonlinear entropy transfer term in the steady state should be negative in the linearly-unstable wavenumber region for the quasilinear model to be valid. 
This speculation is confirmed from the ITG turbulence simulation showing that the quasilinear weights agree well with the nonlinear results in the wavenumber regions where  the nonlinear entropy transfer term is negative. 
It is conjectured from results of this work that energy is generally transferred by turbulence from a particle species with larger entropy production due to particle and heat transport driven by the instability to other species regardless of which species is hotter.
To verify this conjecture, turbulent energy exchange and its predictability in cases of other instabilities such as those driven by trapped electrons and electron temperature gradient remain subjects of future studies.

Furthermore, in principle, the theory and simulation methods for turbulent energy exchange in this study are expected to be applicable or extendable to non-tokamak systems, such as helical systems (stellarators and heliotrons)\cite{Wakatani}.
The study of turbulent energy exchange and quasilinear weights in such general 3D systems 
is also the subject of future research.

\section*{acknowledgements}
The present study is supported in part by the JSPS Grants-in-Aid for Scientific Research Grant No.~19H01879, 24K07000, and in part by the NIFS Collaborative Research Program NIFS23KIPT009.
This work is also supported by JST SPRING, Grant Number JPMJSP2108 and the NINS program of Promoting Research by Networking among Institutions, Grant Number 01422301.
Simulations in this work were performed on “Plasma Simulator” (NEC SX-Aurora TSUBASA) of NIFS with the support and under the auspices of the NIFS Collaboration Research program (NIFS23KIPT009). 

\section*{author declarations}
\subsection*{Conflict of Interest}
The authors have no conflicts to disclose.

\subsection*{Author Contributions}
\noindent
{\bf Tetsuji Kato}: 
Conceptualization (equal);
Data curation (lead);
Formal analysis (equal);
Investigation (lead);
Methodology (equal);
Writing – original draft (lead);
Writing – review \& editing (lead).
{\bf Hideo Sugama}: 
Conceptualization (equal);
Formal analysis (equal);
Investigation (supporting);
Methodology (equal);
Supervision (lead);
Writing – original draft (supporting);
Writing – review \& editing (supporting).
{\bf Tomohiko Watanabe}: 
Methodology (equal);
Writing – review \& editing (supporting). 
{\bf Masanori Nunami}: 
Project administration (lead);
Writing – review \& editing (supporting).

\section*{Data availability}
The data that support the findings of this study are available from the corresponding authors upon reasonable request.

\appendix

\section{Discussion of the effect of electric field}\label{app: shear}
In this work, we follow the local gyrokinetic theory based on the WKB (or ballooning) representation\cite{Antonsen}, and assume that background $E\times B$ velocity is on the order of $(\rho_{ta}/L) v_{ta} =\delta v_{ta}$ (the same order as that of the diamagnetic velocity $\bm{v}_{*a}$) and that the gradient scale length of the background electric field is the same as the equilibrium scale length $L$. 
Under these assumptions, 
the Doppler frequency shift due to the background $E\times B$ velocity 
$\bm{k}_\perp \cdot \bm{v}_E$  is considered to be on the same order as the diamagnetic frequency $\bm{k}_\perp \cdot \bm{v}_{*a}$ 
although the background $E\times B$ velocity shear effect is estimated by 
$\Delta r (\partial v_{E}/\partial r) (\nabla r \times \bm{b})\cdot \nabla \hat{f}_a /|\nabla r| \sim \rho_{ta} (\delta v_{ta}/L)  k_\perp  \hat{f}_a \sim (\delta v_{ta}/L)\hat{f}_a
\sim (\delta k_\perp v_{*a}) \hat{f}_a$, 
the magnitude of which is smaller than terms included in the gyrokinetic equation, Eq.~(\ref{eq:GKequation_nonadiabatic}), by a factor of $\delta$.
Here, $\Delta r$ is the radial correlation length of the fluctuation which is estimated as the typical perpendicular wavelength $1/k_\perp \sim \rho_{ta}$.

If the background $E\times B$ flow is larger and/or the gradient scale length of the background electric field is smaller than assumed above,
the background $E\times B$ flow shear effect cannot be neglected and 
both neoclassical and turbulent processes of transport and energy exchange are considered 
to have $E_r$-dependence different from the results in this paper. 
Also, the coupling of neoclassical and turbulent processes which influences the $E_r$-dependence 
may occur in a global model including the sheared $E_r$ profile.
A nonlinear gyrokinetic equation with large flow velocities on the order of the ion thermal speed is derived in Ref.~\cite{Sugama1998}.

\begin{comment}
The ordering between the perturbed and equilibrium electrostatic potentials is assumed to be  $\phi/\Phi \sim \delta$. 
%
When we suppose that the equilibrium varies over some characteristic length scale $L$ much larger than a thermal gyroradius $\rho_{ta}$ which is assumed the scale length of perturbations, the ordering for radial electric fields is given by
\begin{equation}
    \frac{|\hat{\bm{E}_r}|}{|\bm{E}_r|}= 
    \frac{|\partial\hat{\phi}/\partial r|}{|\partial{\Phi}/\partial r|}\sim 
    \frac{\hat{\phi}/\rho_{ti}}{\Phi/L} \sim 1 
\end{equation}
where the subscript $r$ denotes a radial component. 
%
Since the equilibrium and perturbed electric fields are the same order, both of them appear in Eq.~(\ref{eq:GKequation_nonadiabatic}). 
%
On the other hand, considering the effect of the electric field shear in the flux tube in the same way above, we obtain
\begin{equation}
    \frac{|(\partial^2\Phi/\partial r^2)\rho_{ti}|}{|\partial{\Phi}/\partial r|}\sim 
    \frac{(\Phi/L^2)\rho_{ti}}{\Phi/L} \sim \delta.
\end{equation}
This means that the effect of electric field shear is negligible in local gyrokinetic equations compared with that of electric fields. 
%
Our research follows this assumption.
%
$\Delta r (\partial v_ExB/\partial r) (\partial h / \partial r)  \sim ( \rho / r )(v_t/L)  (k_r \rho)  \hat{f} $,

%
However, the ordering on electric field shear is broken when the gradient of the electrostatic potential is strongly steep as in the H mode pedestal.
%
In this case, the proper gyrokinetics should be used.
\end{comment}

%
Based on the ordering assumption described above, the sum of the last two terms on the right-hand side of Eq.(\ref{eq:transport equation}) 
 can be rewritten as\cite{Sugama1996} 
\begin{eqnarray}
\label{eq:Qturbplus}
& & 
- e_a \Gamma_a^{\rm{turb}} 
    \frac{\partial \Phi}{\partial r}
+ Q^{\rm turb}_a
\nonumber \\ & & 
=  
e_a 
\sum_{\bm{k}_\perp}
\Re \bigg\langle \bigg\langle 
\int d^3v h_{a\bm{k}_\perp}^*
\bigg(
\frac{\partial}{\partial t}
+ i \bm{k}_\perp \cdot \bm{v}_E
\bigg)
\psi_{a\bm{k}_\perp}
 \bigg\rangle  \bigg\rangle
\nonumber \\ & & 
=  
e_a 
\sum_{\bm{k}_\perp}
\Re \bigg\langle \bigg\langle 
\int d^3v 
\bigl[
e^{i \bm{k}_\perp \cdot \bm{v}_E}
h_{a\bm{k}_\perp}
\big]^*
\frac{\partial}{\partial t}
\big[
e^{i \bm{k}_\perp \cdot \bm{v}_E}
\psi_{a\bm{k}_\perp}
\big]
 \bigg\rangle  \bigg\rangle
.\nonumber \\
%\hspace*{7mm}
\end{eqnarray}
As pointed out in Ref.\cite{Sugama1996}, 
the background radial electric field $E_r = -\partial \Phi/\partial r$
enters the gyrokinetic equation, Eq.~(\ref{eq:GKequation_nonadiabatic}) only 
in the form of the Doppler shift 
$(\partial / \partial t + i \bm{k}_\perp \cdot \bm{v}_E)$ 
and does not appear explicitly in Eqs.~(\ref{eq: NonAdGK})--(\ref{eq:B_para}).
Therefore, for the solutions of Eq.~(\ref{eq:GKequation_nonadiabatic}) and Eqs.~(\ref{eq:Poisson})--(\ref{eq:B_para}), 
$e^{i \bm{k}_\perp \cdot \bm{v}_E} h_{a\bm{k}_\perp}$
and 
$e^{i \bm{k}_\perp \cdot \bm{v}_E} \psi_{a\bm{k}_\perp}$
are independent of $E_r$. 
Then, 
even though each of the first and second terms on the left-hand side of 
Eq.~(\ref{eq:Qturbplus}) depends on $E_r$, 
their sum does not. 
In the present paper, 
we evaluate $Q^{\rm turb}_a$ by the gyrokinetic simulation for $E_r = 0$ 
although 
we should note that 
the results of $Q^{\rm turb}_a$ for $E_r = 0$ are equivalent to 
those of $e_a \Gamma_a^{\rm{turb}} E_r + Q^{\rm turb}_a$ for any $E_r$. 
We also find from Eq.~(\ref{eq:gradients}) that the right-hand side of the entropy balance equation, 
Eq.~(\ref{eq:EBequation_h}), contains 
this sum of the terms 
$e_a \Gamma_a^{\rm{turb}} E_r + Q^{\rm turb}_a$
which takes a value independent of $E_r$.

\section{Derivation of Eq.~(\ref{eq:TEE_parts})}\label{app: derivation}

In this Appendix~\label{app: derivation}, the derivation of Eq.~(\ref{eq:TEE_parts}) is presented.
We first note 

\begin{eqnarray}\label{eq: Leibniz}
\Bigg\langle\Bigg\langle e_a\int d^3v h_{a\bm{k}_\perp}^*\frac{\partial \psi_{a\bm{k}_\perp}}{\partial t}\Bigg\rangle\Bigg\rangle
&=&
\frac{\partial }{\partial t}
\Bigg\langle\Bigg\langle e_a\int d^3v \psi_{a\bm{k}_\perp}h^*_{a\bm{k}_\perp} \Bigg\rangle\Bigg\rangle
\nonumber \\
& & -
\Bigg\langle\Bigg\langle e_a\int d^3v \psi_{a\bm{k}_\perp}\frac{\partial h^*_{a\bm{k}_\perp}}{\partial t} \Bigg\rangle\Bigg\rangle,\nonumber \\
\end{eqnarray}
where $\langle \langle \cdots \rangle \rangle$ denotes a double average over the ensemble and the flux surface.
The first term of the right-hand side in Eq.~(\ref{eq: Leibniz}) is negligible because of the time transport scale ordering $\left( \partial \langle \langle \cdots \rangle \rangle / \partial t\right)/\langle \langle \cdots \rangle \rangle = \mathcal{O}(\delta^2)$. 
Substituting Eq.~(\ref{eq:GKequation_nonadiabatic}) into Eq.~(\ref{eq: Leibniz}), we obtain
\begin{eqnarray}\label{eq: TEEwithGKequation}
&&\Bigg\langle\Bigg\langle -e_a\sum_{\bm{k}_\perp}\int d^3v \psi_{a\bm{k}_\perp}\frac{\partial h^*_{a\bm{k}_\perp}}{\partial t}\Bigg\rangle\Bigg\rangle \nonumber \\
&&=\Bigg\langle\Bigg\langle e_a\sum_{\bm{k}_\perp}\int d^3v \psi_{a\bm{k}_\perp}\bigg( v_{\parallel } \bm{b}\cdot\nabla h^*_{a\bm{k}_\perp} + i\bm{k}_\perp \cdot \bm{v}_{da} h^*_{a\bm{k}_\perp} \nonumber \\
& &\hspace{10pt}-\frac{c}{B} \sum_{\bm{k}_\perp'+\bm{k}_\perp''=\bm{k}_\perp}\left[ \bm{b}\cdot\left( \bm{k}_\perp'\times\bm{k}_\perp'' \right)\right] \psi^*_{a\bm{k}_\perp'}h^*_{a\bm{k}_\perp''}  -C_a^{GK*} \bigg)\Bigg\rangle\Bigg\rangle. \nonumber \\
\end{eqnarray}
On the right-hand side of Eq.~(\ref{eq: TEEwithGKequation}), a pure imaginary term disappears and a time derivative term is neglected again because of the transport time scale ordering.
We immediately see that Eqs.(\ref{eq:perp_heating}) and (\ref{eq:collision_psi}) corresponds to the second and fourth terms on the right-hand side of Eq.(\ref{eq: TEEwithGKequation}).
Equation~(\ref{eq:psi_heating}) is also derived from the third term by $\sum_{\bm{k}_\perp'+\bm{k}_\perp''=\bm{k}_\perp}\left[ \bm{b}\cdot\left( \bm{k}_\perp'\times\bm{k}_\perp'' \right)\right]= \sum_{\bm{k}_\perp'+\bm{k}_\perp''=\bm{k}_\perp}\left[ \bm{b}\cdot\left( \bm{k}_\perp'\times\bm{k}_\perp \right)\right]$.
When phase space variables ($\bm{x}$, $w=m_av^2/2$, $\mu=m_av_\perp^2/2B$) are used, the integral over velocity space is described as
\begin{equation}
    \int d^3v=\frac{B}{m_a^2}\sum_\sigma\oint d\xi \int_0^\infty dw \int_0^{w/B}d\mu \frac{1}{|v_\parallel|},
\end{equation}
where $\xi$ is the gyrophase and $\sigma=v_\parallel/|v_\parallel|=\pm1$.
The first term can be written as
\begin{eqnarray}\label{eqB: paralllel heating }
&& \Bigg\langle\Bigg\langle e_a \int d^3v \psi_{a\bm{k}_\perp} v_{\parallel } \bm{b}\cdot\nabla h^*_{a\bm{k}_\perp} \Bigg\rangle\Bigg\rangle \nonumber \\
    &&=\Bigg\langle\Bigg\langle \frac{e_a}{m_a^2}\sum_\sigma\oint d\xi \int_0^\infty dw \int_0^{w/B}d\mu \bigg( \bm{B}\cdot \nabla \big(\sigma h^*_{a\bm{k}_\perp} \psi_{a\bm{k}_\perp}\big)\bigg)\Bigg\rangle\Bigg\rangle \nonumber \\
    &&\hspace{10pt} -\Bigg\langle\Bigg\langle e_a \int d^3v \big(h^*_{a\bm{k}_\perp} v_{\parallel } \bm{b}\cdot\nabla \psi_{a\bm{k}_\perp}\big)\Bigg\rangle\Bigg\rangle \nonumber \\
    &&=\Bigg\langle\Bigg\langle \frac{e_a}{m_a^2}\left(\bm{B}\cdot \nabla\right) \left( \oint d\xi \int_0^\infty dw \int_0^{w/B}d\mu \left(\sum_\sigma\sigma h^*_{a\bm{k}_\perp} \psi_{a\bm{k}_\perp}\right)\right)\Bigg\rangle\Bigg\rangle \nonumber \\
    &&\hspace{10pt} -\Bigg\langle\Bigg\langle e_a \int d^3v \big(h^*_{a\bm{k}_\perp} v_{\parallel } \bm{b}\cdot\nabla \psi_{a\bm{k}_\perp}\big)\Bigg\rangle\Bigg\rangle.
\end{eqnarray}
Since the integral range of $\mu$ depends on position $\bm{x}$, the boundary of the integral range needs to be taken into account when moving $\nabla$ outside of the integral on the right-hand side of Eq.(\ref{eqB: paralllel heating }).
However, when $\mu =w/B$, it is a bounce point($v_{\parallel} = 0$) and the integral function $\sum _\sigma \sigma h^*_{a\bm{k}_\perp} \psi_{a\bm{k}_\perp}$ at the point is zero.
Therefore, the boundary effect does not affect the integral calculations.
The flux surface average of the first term of the right-hand side in Eq.~(\ref{eqB: paralllel heating }) vanishes due to $\langle \bm{B}\cdot\nabla \cdots \rangle =0$.

Then, we obtain 
\begin{eqnarray}\label{eqB: paralllel heating2 }
& &\Bigg\langle\Bigg\langle  e_a \int d^3v \psi_{a\bm{k}_\perp} v_{\parallel } \bm{b}\cdot\nabla h^*_{a\bm{k}_\perp}\Bigg\rangle\Bigg\rangle \nonumber \\
& &\hspace{10pt} = \Bigg\langle\Bigg\langle  -e_a \int d^3v h^*_{a\bm{k}_\perp} v_{\parallel } \bm{b}\cdot\nabla \psi_{a\bm{k}_\perp}\Bigg\rangle\Bigg\rangle,
\end{eqnarray}
which gives Eq.~(\ref{eq: parallel_heating}).

\section*{references}

\end{document}